 \documentclass[useAMS,usenatbib]{mn2e}
\usepackage{psfig}
\usepackage{natbib}
\usepackage{amssymb}
\usepackage[utf8]{inputenc}

\newcommand{\be}{\begin{equation}}
\newcommand{\en}{\end{equation}}

\def\zabs{$z_{\rm abs}$}
\def\zem{$z_{\rm em}$}

\def\mgii{Mg~{\sc ii}}

\def\caii{Ca~{\sc ii}}
\def\mnii{Mn~{\sc ii}}
\def\ni2{Ni~{\sc ii}}
\def\mgi{Mg~{\sc i}}
\def\alii{Al~{\sc ii}}
\def\aliii{Al~{\sc iii}}
\def\civ{C~{\sc iv}}
\def\siiv{Si~{\sc iv}}

\def\si2{Si~{\sc ii}}
\def\feii{Fe~{\sc ii}}
\def\h1{H~{\sc i}}
\def\feiii{Fe~{\sc iii}}

\def\feiia{Fe~{\sc ii}$\lambda$2600}

\def\kms{km~s$^{-1}$}
\title[Time variability of FeLoBALs]{Probing the time variability of five Fe low broad absorption line quasars \thanks{Uses archival data based on observations carried out at the European Southern Observatory (ESO), under programmes
267.B-5698 and 71.B-0121.}}
\author[M. Vivek et al.]{M. Vivek$^{1}$\thanks{E-mail:vivekm@iucaa.ernet.in}, R. Srianand$^{2}$, P. Petitjean$^{3}$, P. Noterdaeme$^{3}$, V.Mohan$^{2}$, \newauthor A. Mahabal$^{4}$ \& V.C. Kuriakose$^{1}$\\
$^{1}$Department of Physics, Cochin University of Science and Technology, Kochi 682022, India\\
$^{2}$Inter University Centre for Astronomy and Astrophysics, Pune 410007, India\\
$^{3}$UPMC-CNRS, UMR7095,  Institut $d '$Astrophysique de Paris, 98bis Boulevard Arago, 75014 Paris, France\\ 
$^{4}$Caltech, MC 249-17, Pasadena, CA 91125, USA }
\begin{document}
\date{Accepted . Received ; in original form }
\pagerange{\pageref{firstpage}--\pageref{lastpage}} \pubyear{2002}
\maketitle

%%%%%%%%%%%%%%%%%%%%%%%%%%%%%%%%%%%%%%%%%%%%%%%%%%%%%%%%%%%%%%%%%
\label{firstpage}
\begin{abstract}
We study the time variability of five Fe Low ionization  Broad Absorption 
Line (FeLoBAL) QSOs using repeated spectroscopic observations with the 
2m telescope at IUCAA Girawali observatory (IGO) spanning  an interval of 
upto 10 years. We report a dramatic variation in  Al~{\sc iii} and 
Fe~{\sc iii} fine-structure lines  in the spectra of  
SDSS J221511.93-004549.9 ($z_{\rm em} \sim$ 1.478). However, there is no such 
strong variability shown by the \civ\ absorption.  This source is known to be 
unusual with  (i) the continuum emission dominated
by Fe emission lines, (ii) Fe~{\sc iii} absorption being stronger than 
Fe~{\sc ii} and (iii) the apparent ratio of Fe~{\sc iii} UV 48 to Fe~{\sc iii} UV 34 
absorption suggesting an inverted population ratio.   
This is the first reported detection of time variability in the  
Fe~{\sc iii} fine-structure lines in QSO spectra. There is a strong
reduction in the absorption strength of these lines between year 2000
and 2008. 
Using the template fitting techniques, we show that the apparent inversion 
of strength of UV lines could be related to the complex spectral energy 
distribution of this QSO.  The observed  variability can be related to 
change in the ionization state of the gas  or due to transverse motion of 
this absorbing gas. 
The shortest variability timescale of \aliii\ line gives a  
lower limit on the electron density of the absorbing gas as 
$n_e \geq 1.1 \times 10^4$ cm$^{-3}$.  
The remaining 4 FeLoBALs do not show any  changes beyond the measurement uncertainties either in optical depth or in the velocity structure.  We present the  long-term photometric light curve for all of
our sources. Among them only SDSS  J221511.93-004549.9 shows significant ($\geq$ 0.2 mag) variability.
\end{abstract}


\section{Introduction}

 Broad Absorption Line (BAL) quasars form $\sim$ 20 - 40\% of the QSO 
population \citep{hewett03,reichard03,trump06,Dai08,Stalin11,allen11} 
and are characterized by the presence of broad absorption lines spreading 
over 5000 - 50000 km s$^{-1}$ \citep{green01}. BAL QSOs are further 
classified into three sub-groups based on the material producing the 
BAL troughs. High-ionization BALs(HiBALs) contain strong, broad absorption 
troughs of highly ionized species such as \civ, O~{\sc vi} and N~{\sc v} 
and are typically identified through the presence of \civ\ absorption 
troughs. Low-ionization BALs (LoBALs) contain absorption from 
low-ionization species such as Mg~{\sc ii}, \alii\ or \aliii\ in 
addition to the standard absorption lines seen in HiBALs. The 
LoBALs comprise about 15\% of the BAL population.  A small 
sub-set of LoBALs with excited-state \feii\ or \feiii\ absorption 
are termed as FeLoBALs \citep{wampler95}. Only 13$\%$ of the LoBALs 
(i.e 1\% of the total BAL population)  are FeLoBALs.
The catalog of BAL quasars in SDSS-DR3 by \citet{trump06} has 138 FeLoBALs. 
As these QSOs are very rare, they are not a well studied population of 
BALQSOs. 

\par
FeLoBALs are also found to be the most reddened  
ones among BAL quasars \citep{reichard03}.  
The FeLoBALs are found to be  extremely IR-luminous \citep{farrah07},
with IR luminosities comparable to those of 
Ultra-Luminous Infrared galaxies (ULIRG). 
Hence, FeLoBAL phenomenon could be considered as a transition stage 
in a ULIRG when the star burst is at or near its end, and the central 
QSO is starting to throw off its dust cocoon \citep{Voit93,farrah07}. 
This scenario strongly favours the theory of BAL QSO to be an evolutionary 
stage in a quasar lifetime, rather than an orientational effect. 
The \civ\ BALs are referred to be located at a distance $\lesssim$ 1 pc 
\citep[e.g.][]{capellupo11}. However, ionization parameter and density 
estimates of some of the FeLoBALs are consistent  with the absorbing gas  
at $\geq$ 1 kpc from the continuum source \citep{korista08,moe09,dunn10,bautista10,faucher11}. This together 
with very small inferred thickness of the cloud prompted 
Faucher-Gigu{\`e}re et al. to suggest that FeLoBALs must be formed  in situ at 
large radii through interaction  of QSO blast wave in the interstellar 
medium (ISM) of the host galaxy. In this scenario the time evolution of 
the post shock gas will be seen as the absorption line variability. 
%++++++++++++++++++++++++++++++++++++++++++++++++++++++++++++++++++++++++++++++++++++++++
 \begin{table*}
% \begin{sidewaystable}[ht]
\caption{Log of observations}
\flushbottom
\begin{tabular}{|c|c|c|c|c|c|c|c|}
\hline
QSO&Instrument& Date&MJD&Exposure Time&$\lambda$ Coverage&Resolution&S/N$^a$\\
&&&&(mins)&($\AA$)&(kms$^{-1}$)&\\
\hline
&SDSS&29-09-2000 &51816&45x1 &3800-9200&150&44\\
&SDSS& 29-11-2000&51877 &50x1 &3800-9200&150&40\\
J0300$+$0048&SDSS& 23-10-2001&52205 &45x1 &3800-9200&150&33\\
&IGO/IFOSC 7&12-12-2007&54446&35x2&3800-6840&310&5\\
&IGO/IFOSC 7&27-12-2008&54817&45x8&3800-6840&310&54\\
\hline
&SDSS&15-01-2001 &51924&80x1 &3800-9200&150&27\\

&IGO/IFOSC 8&14-12-2007&54448&45x3&5800-8350&240&10\\
&IGO/IFOSC 7&09-01-2008&54474&45x3&3800-6840&310&23\\

J0318$-$0600&IGO/IFOSC 8&27-11-2008&54797&45x3&5800-8350&240&21\\
&IGO/IFOSC 7&31-01-2009&54862&45x4&3800-6840&310&19\\

&IGO/IFOSC 8&23-12-2009&55188&45x4&5800-8350&240&19\\
&IGO/IFOSC 7&24-12-2009&55189&45x4&3800-6840&310&17\\
\hline
&SDSS&19-11-2001&52232&48x1&3800-9200&150&26\\
J0835$+$4242&IGO/IFORS 1&14-12-2007&54448&45x3&3270-6160&370&9\\
&IGO/IFOSC 7&05-12-2008&54805&45x3&3800-6840&310&21\\
&IGO/IFOSC 7&21-01-2010&55217&45x3&3800-6840&310&29\\        
\hline
&SDSS&15-02-2002&52320&50x1&3800-9200&150&40\\
&IGO/IFOSC 7&20-12-2006&54089&45x3&3800-6840&310&8\\
J0840$+$3633&IGO/IFOSC 7&12-12-2007&54446&45x3&3800-6840&310&23\\
&IGO/IFOSC 7&07-12-2008&54807&45x4&3800-6840&310&26\\
&IGO/IFOSC 7&17-12-2009&55182&45x6&3800-6840&310&50\\
\hline
&SDSS&04-09-2000&51804&53x1&3800-9200&150&41\\
&VLT/UVES&11-08-2001&52145&60x2&3000-11000&7.3&43\\
&VLT/FORS1&20-09-2003&52902&5x1&3300-11000&250&62\\
&IGO/IFOSC 7&31-10-2008&54783&45x3&3800-6840&310&30\\
&IGO/IFOSC 7&01-11-2008&54784&45x1&3800-6840&310&23\\
J2215$-$0045&IGO/IFOSC 8&01-11-2008&54784&45x4&5800-8350&240&39\\
&IGO/IFORS 1&02-01-2010&55211&45x1&3270-6160&370&11\\
&IGO/IFORS 1&04-01-2010&55213&45x3&3270-6160&370&15\\
&MagE&10-08-2010&55431&15x1&3100-10000&70&16\\
&IGO/IFORS 7&14-12-2010&55557&45x2&3800-6840&310& 7\\
&IGO/IFORS 7&15-12-2010&55558&45x1&3800-6840&310&9 \\
&IGO/IFORS 7&18-12-2010&55561&45x2&3800-6840&310&11 \\
&IGO/IFORS 1&12-12-2011&55920&45x3&3270-6160&370&14\\
&IGO/IFORS 1&14-12-2011&55922&45x4&3270-6160&370&17\\

\hline
\end{tabular}
 \begin{flushleft}
  $^a$ calculated  over the wavelength range 5800-6200$\AA$ \\
      \end{flushleft}
 \label{log}
\end{table*}
%+++++++++++++++++++++++++++++++++++++++++++++++++++++++++++++++++++++++++++++++++++

\par
Repeated BAL monitoring studies could greatly help us in understanding 
the location and physical conditions in the absorbing gas and the physical 
mechanisms responsible for quasar outflows. Time variability of \siiv\ and 
\civ\ BALs in individual sources has been reported in several BALQSOs: 
Q1303+308 \citep{foltz87}, Q1413+113 \citep{turnshek88}, 
Q1246-057 \citep{smith88}, UM 232 \citep{barlow89}, 
QSO CSO 203 \citep{barlow92}, Tol 1037-270 \citep{anand01},
J1054+0348 \citep{hamann08}
and FBQS J1408+3054 \citep{hall11}. 
\citet{lundgren07} have reported  significant time variability among a 
sample of 36 \civ\ BALs on rest frame timescales shorter than 1 year. 
\citet{gibson08} also report a similar result in a sample of 13 quasars
(1.7 $\leq z \leq$ 2.8) over 3-6 (rest frame) years.  \citet{gibson10} 
investigated the \civ\ BAL variability of 14 sources at redshifts 
$>$ 2.1 and report complex variations in the sample.  In these cases, 
change in the rest equivalent width of absorption lines are used as an 
indicator of variability. This requires continuum fitting that is in 
general ambiguous in the case of BALQSOs. \citet{capellupo11} have studied 
\civ\ BAL variability in 34 luminous QSOs over short (4-9 months) and 
long (3.8-7.7 years) timescales. They use flux differences between two 
epoch data at the absorption trough to quantify the variability.
However, no such studies are
reported for the LoBALs and in particular there is no case reported
showing the variability of absorption lines originating from excited 
fine-structure levels in the case of BALQSOs.

QSO feedback plays an important role in the evolution of the host galaxy. 
If FeLoBALs represent a transition state (or post shock ISM gas) as 
discussed above, then repeated spectroscopic monitoring of these sources 
will shed light on the evolutionary scenarios  of QSOs. In this paper, 
we report the nature of  time variability in a sample of five  FeLoBAL 
quasars. This is  a sub-sample of our ongoing spectroscopic monitoring 
campaign on time variability of absorption lines in a sample of 27 LoBAL 
quasars. Three of the sources, SDSS J030000.57+004828.0 (hereafter SDSS 
J0300+0048), SDSS J031856.62-060037.7 (hereafter SDSS J0318-0600) and 
SDSS J221511.93-004549.9 (hereafter SDSS J2215-0045) are from the 23  
Sloan Digital Sky Survey (SDSS) Early Data Release (EDR) listed quasars 
identified by \citet{hall02} as BAL quasars exhibiting various unusual 
properties.  The other two sources are SDSS J083522.77+424258.3 
(hereafter SDSS J0835+4242) and SDSS J084044.41+363327.8 (hereafter SDSS 
J0840+3633). 

This manuscript is arranged as follows.
In section 2 we provide details of observation and data reduction.
Discussion on individual sources in our sample are presented in
Section 3. Implications of the variability seen in  SDSS J2215-0045 are
presented in section 4 and the main results are summarized in Section 5.

\section{Observation and Data Reduction}
Most of the new observations presented here were carried out using the 
2m telescope at IUCAA Girawali observatory (IGO).
The spectra were obtained using the IUCAA Faint 
Object Spectrograph (IFOSC).  The detailed log of these observations together 
with that of the archival SDSS data and the data from the literature are 
given in Table~\ref{log}.  Spectra were originally obtained mainly 
using three grisms, Grism 1, Grism 7 and Grism 8 of IFOSC in combination  
with 1.5 arcsec slit. This  combinations has a wavelength coverage between  
3270 - 6160~\AA, 3800 - 6840~\AA\ and 5800 - 8350~\AA\ 
for the above three grisms respectively. Typically the observations 
were split in to exposures of 45 minutes. All the raw frames were 
processed using standard IRAF{\footnote {IRAF is distributed by the 
National Optical Astronomy Observatories, which are operated by the 
Association of Universities for Research in Astronomy, Inc., 
under cooperative agreement with the National Science Foundation.}} 
tasks. One dimensional spectra were extracted from the frames using the  
``doslit'' task in IRAF. We opted for the variance-weighted extraction 
as compared to the  normal one. Wavelength calibrations were done 
using standard helium neon lamp spectra and flux calibrations were done 
using a standard star spectrum observed on the same night. Air-to-vacuum 
conversion was applied before adding the spectra. Individual spectra were 
combined using 1/$\sigma^2$ weighting in each pixel after scaling the 
overall individual spectra to a common flux level within a sliding 
window. The error spectrum was computed taking into account proper 
error propagation during the combining process.

Our IGO spectrum of J2215-0045 taken in 2008 has shown dramatic variability 
as compared to the archival SDSS spectrum. To confirm this trend we obtained 
a moderate resolution echellete spectrum of the source using the  
Magellan Echellette (MagE) Spectrograph mounted on the Clay (Magellan II) 
telescope. We used 1 arcsec slit and 1$\times$1 binning for our observations. 
The available grating in combination with the 1 arcsec slit gives a 
resolution of R=4200 and wavelength coverage of 3000-10000 $\AA$. 
Spectrophotometric standards were also observed for flux calibrations. 
MagE Data was reduced using the MagE Spectral Extractor (MASE) pipeline 
\citep{bochanski09}. MASE is an IDL based pipeline containing a graphic 
user interface (GUI) for reducing MagE data. VLT-UVES  data 
(Program ID: 267.B-5698) taken on 2001  is available for this source 
in the VLT archive. We also used the  VLT-FORS1 data  
(Program ID: 71.B-0121) obtained for this source in 2003 and made 
available to us by Dr. DiPompeo \citep[][]{dipompeo11}.

\begin{figure*}
 \centering
 \psfig{figure=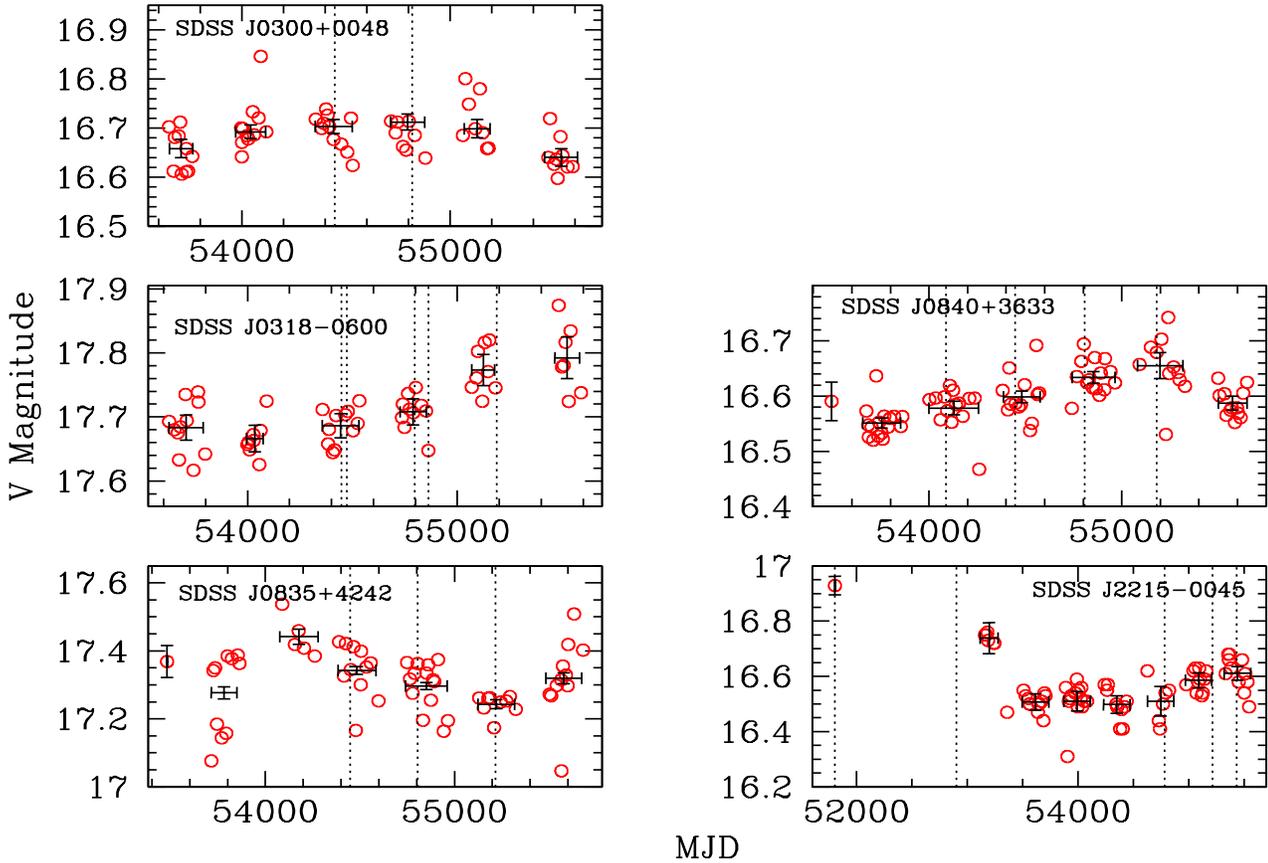,width=1.0\linewidth,height=0.7\linewidth,angle=270}
 \caption{Light curves of five FeLoBAL quasars in the sample.  
The  dotted vertical lines mark the epochs (MJDs) of spectroscopic 
observations. The average magnitudes obtained for closely spaced 
observations are overplotted as points with error bars for all sources. 
The x axis error bars corresponds to the time range over which the 
magnitudes are averaged.}
\label{fig5}
 \end{figure*}

  In  Fig.~\ref{fig5},  we plot for all our sources Johnson's V
magnitude  from the Catalina Real-Time Transient Survey (CRTS; \citet{Drake09}) as described below.  
CRTS operates with an unfiltered set up and the resulting magnitudes are converted to V magnitudes
using the transformation  equation V=V$_{ins}$+a(v)+b(v)*(B-V), where, $V_{ins}$ is the  observed open magnitude, a(v) and  b(v) are the zero point and the slope. The zero point and slope are obtained from three or more comparison stars in the same field with the zero point typically being of the order of 0.08. CRTS provides four such observations taken 10 minutes apart on a given night. For our lightcurves we have averaged these four points (or less if one or more of those coincided with bad areas) and plotted those against the modified Julian date (MJD).
The data used in  these light curves are taken between 
April 2005 to July 2010. The dotted vertical lines show the epochs of our
spectroscopic observations.  For, J2215-0045, we could get the photometric data
since 2004. The first point in the light curve for this source is obtained
using $g$ and $r$ magnitudes in the SDSS and the transformation equation
given by \citet{jester05} for QSOs at $z\le 2.1$. For the other three
sources we did not transform the SDSS magnitudes to Johnson V magnitude as 
$g$ and $r$ band fluxes are affected by strong broad absorption lines. 
For all the  sources, we have overplotted the average magnitudes obtained for closely spaced observations. 
The period over which the 
magnitudes are averaged are shown by x-axis error bars.

From the light curves, it is clear that continuum variability is 
apparent in all the sources. However, continuum variability at the 
level of $>$ 0.2 mag is seen only in the case of J2215-0045. 
The source has brightened by 0.3 mag when we consider only the CRTS 
points. The brightening could be up to 0.5 mag when we also include 
the transformed SDSS mag.

%+++++++++++++++++++++++++++++++++++++++++++++++++++++++++++++++++++++++++++++++++++++++++++++++++++++++++++++++
%+++++++++++++++++++++++++++++++++++++++++++++++++++++++++++++++++++++++++++++++++++++++++++++++++++++++++++++++++
%+++++++++++++++++++++++++++++++++++++++++++++++++++++++++++++++++++++++++++++++++++++++++++++++++++++++++++++++++
\section{Notes on Individual Objects}

 In this section, we summarize the properties of individual absorbers in our sample and discuss our  monitoring results. 
\subsection{SDSS J0300+0048}

SDSS J0300+0048 (\zem=0.8918) is  part of a binary QSO with 
SDSS J025959.69+004813.5 (\zem=0.8923), a non-BAL QSO at a projected
separation of 19.8 arc sec. It is only the fourth QSO known to have a 
\caii\ BAL trough. 
The outflow is blue shifted by ~1650 kms$^{-1}$ from the systemic redshift
of the QSO. \citet{hall03a} obtained the high resolution UVES spectrum 
of this source and found that extremely broad \mgii, \feii\ and its 
fine-structure line absorptions are also present along with the strong 
\caii\ absorption. They also found that the lowest velocity BAL 
region has a strong \caii\ absorption without  significant associated 
excited \feii\ absorption, while the higher velocity excited 
\feii\ absorption region has very little \caii\ absorption. 
% Hall et al (2003) studied the VLT-UVES spectrum of this unusual quasar. 
%It is one of most metal rich quasars known with 15 times solar metallicity. 
The reported Ca~{\sc ii}, \mgii\ and \mgi\ column densities are very high and 
the corresponding gas phase metallicity is found to be fifteen times the
solar value. Comparing this large column densities to that of  QSO J2359-1241, 
\citet{hall03a} argued that this source must have a strong hydrogen 
ionization front where the \caii\ exists outside the \h1 front. 
As the lowest ionization gas is found at lowest velocities, they explained 
the detached flow in SDSS J0300+0048 by the radiatively driven disk wind 
model by \citet{murray95}. Apart from the BAL, there is a narrow 
associated absorption system detected by Ca~{\sc ii}, \mgii , \mgi\ 
and  \feii\ lines at a redshift of \zabs=0.8918.
\citet{hall03a} have also suggested that the optical part of the
QSO continuum is dominated by the Fe~{\sc ii} and 
Fe~{\sc iii} emission line blends. An important feature of this spectrum is that the flux values never 
return to the continuum values  \citep[as shown by][]{hall03a} shortward 
of the rest wavelength 3000 \AA\ ($\lambda_{obs} \sim$ 5675 \AA). Also 
emission bumps seen around the rest wavelength range 3000-3300 \AA\ 
($\lambda_{obs} \sim$ 5675 - 6240 \AA ) are consistent with them being
dominated by Fe emission.% (see the top panel of Fig~\ref{fig1} for comparison).
%++++++++++++++++++++++++++++++++++++++++++++++++++++++++++++++++++++++++++++++++++++++++++++++++++++++++++++++
\begin{figure*}
 \centering

\begin{tabular}{c c}
\psfig{figure=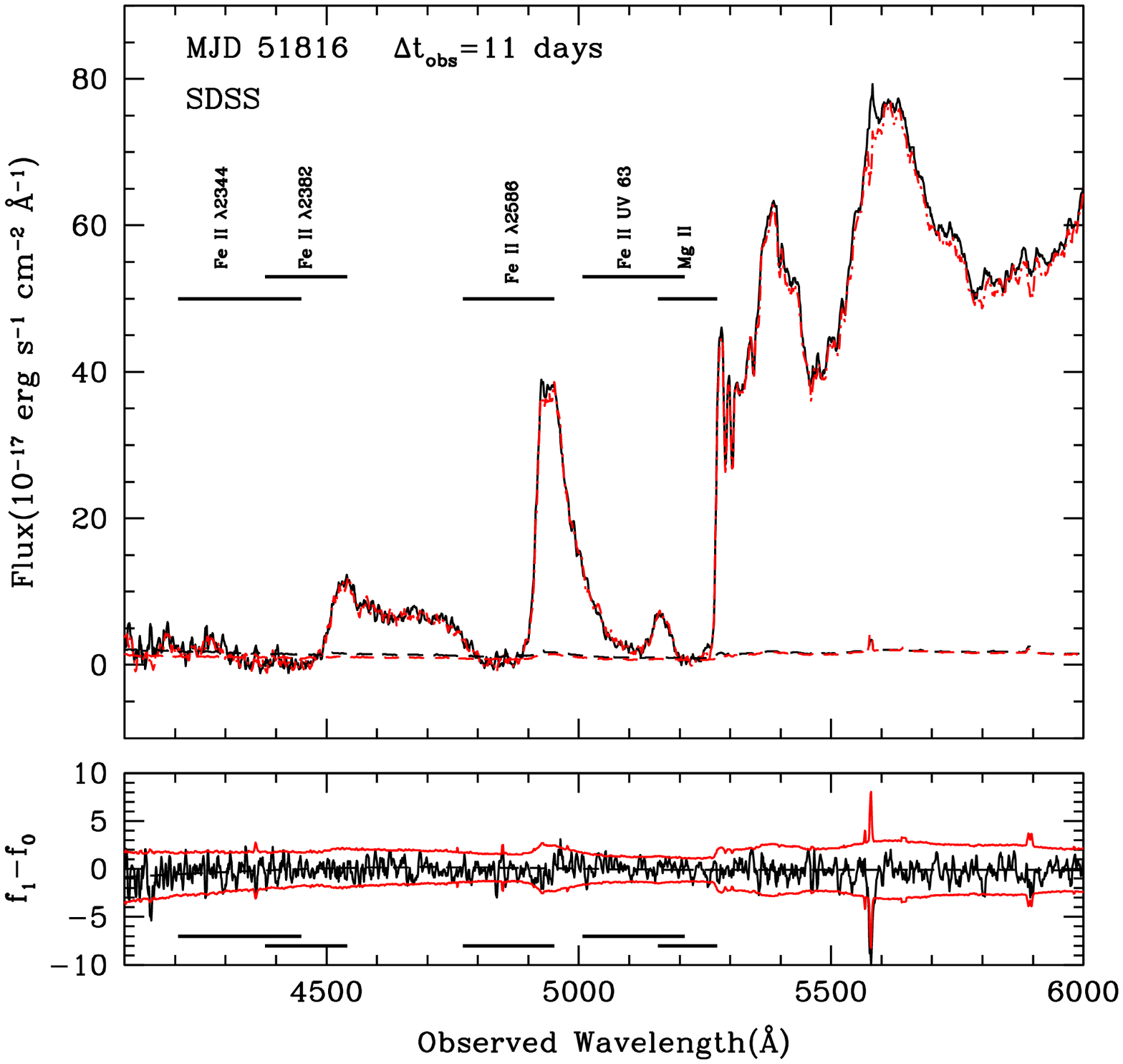,width=9cm,height=6cm}&%,bbllx=21bp,bblly=200bp,bburx=570bp,bbury=705bp,clip=yes}&
\psfig{figure=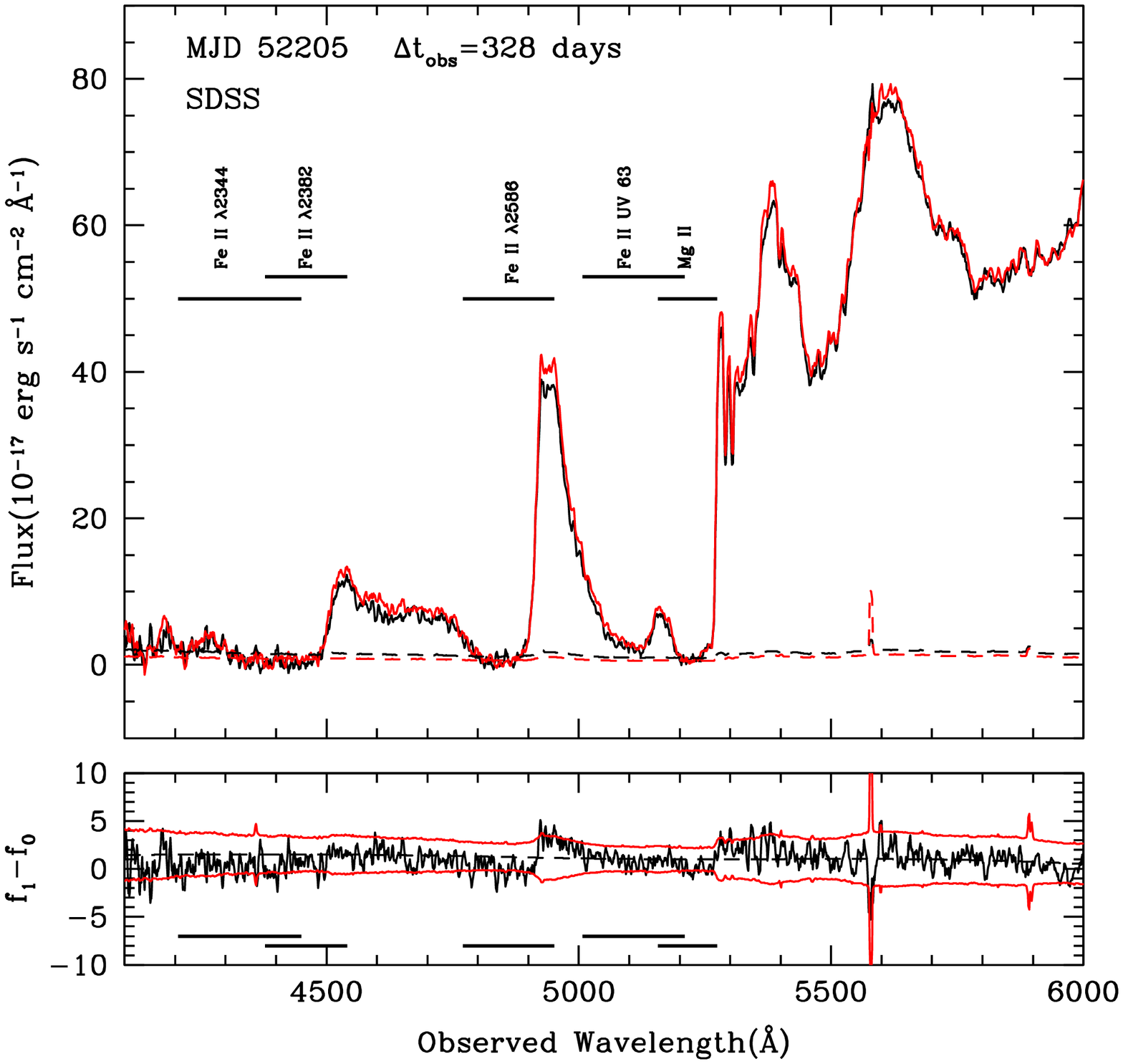,width=9cm,height=6cm}\\%,bbllx=51bp,bblly=200bp,bburx=570bp,bbury=705bp,clip=yes}\\
\psfig{figure=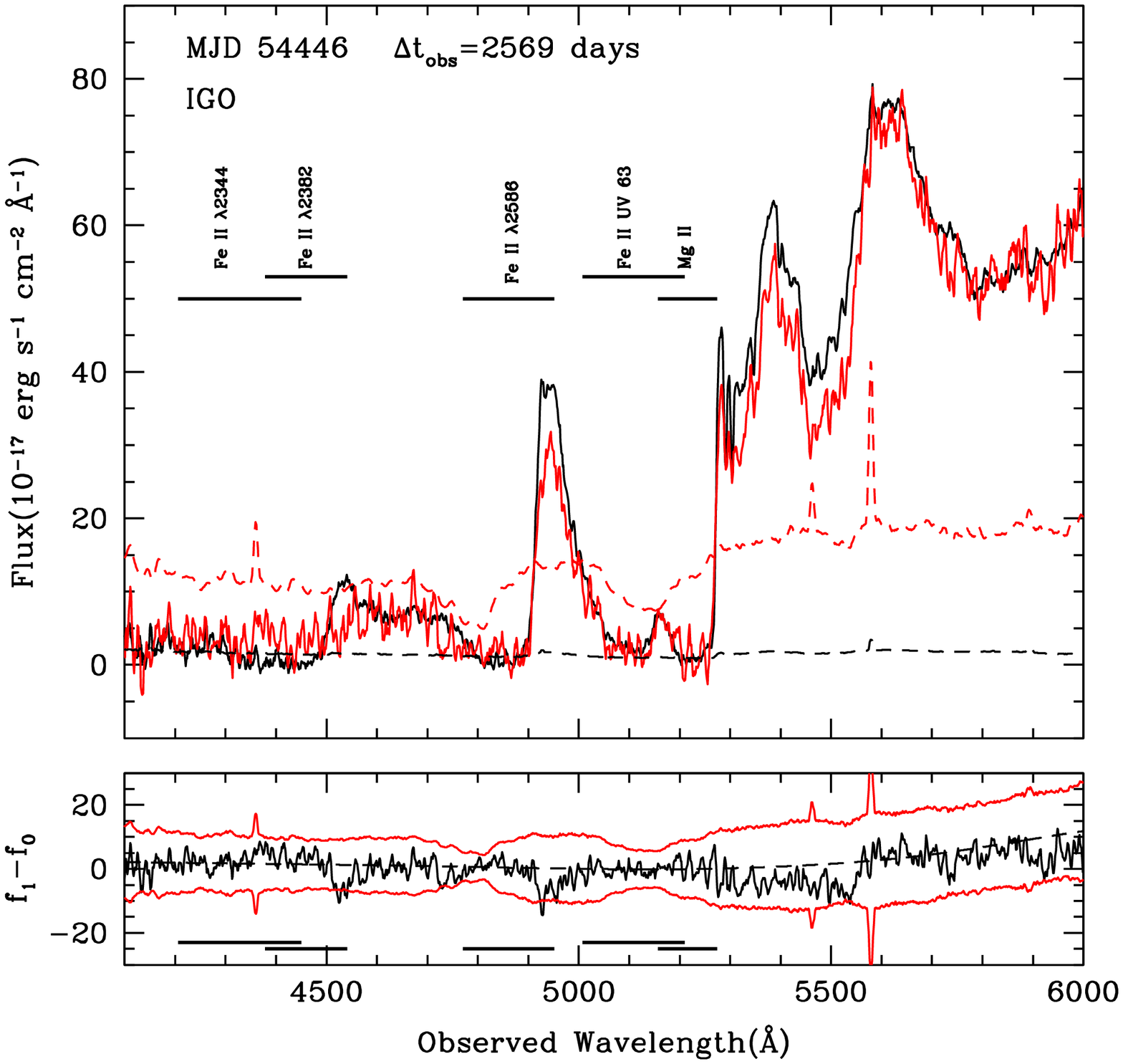,width=9cm,height=6cm}&%,bbllx=21bp,bblly=150bp,bburx=570bp,bbury=698bp,clip=yes}&
\psfig{figure=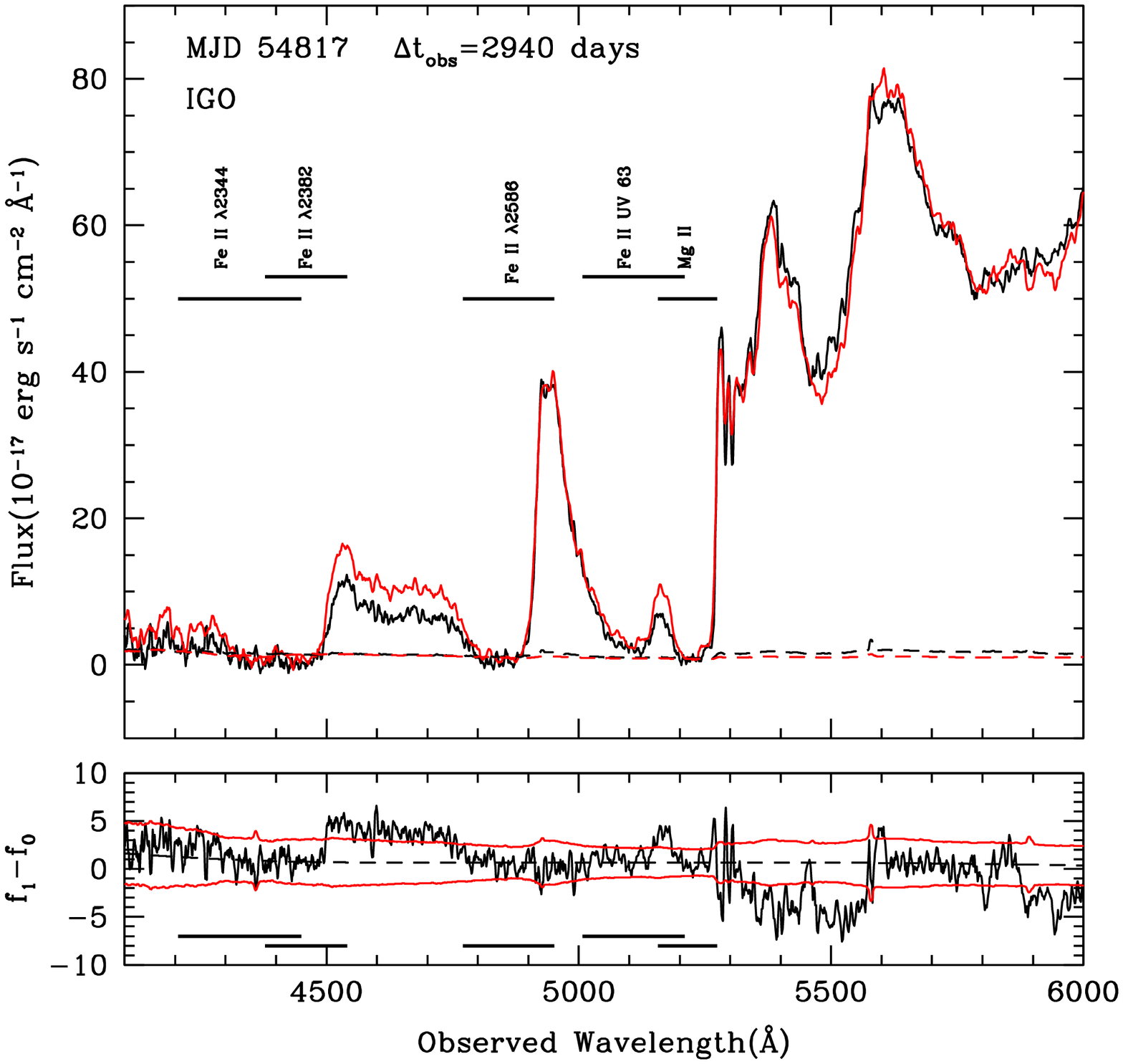,width=9cm,height=6cm}\\%,bbllx=51bp,bblly=200bp,bburx=570bp,bbury=705bp,clip=yes}\\

\end{tabular}
\caption{  Spectra of SDSS J0300+0048 observed in SDSS (observed on
MJD 51816,52205) 
and  IGO (observed on MJD 54446,54817) (in red/gray) are overplotted with the 
reference SDSS spectrum (black) observed on MJD 51877. The flux scale applies 
to the reference SDSS spectrum  and all other spectra are scaled in 
flux to match the reference spectrum. In each plot the error spectra 
are also shown. The difference spectrum for the corresponding 
MJD's are plotted in the lower panel of each plot. 1 $\sigma$ error is plotted 
above and below the mean. The regions of absorption lines  are 
marked  with horizontal lines.  }
\label{fig1}
\end{figure*}

\begin{figure*}
 \centering
\begin{tabular}{c c}
\psfig{figure=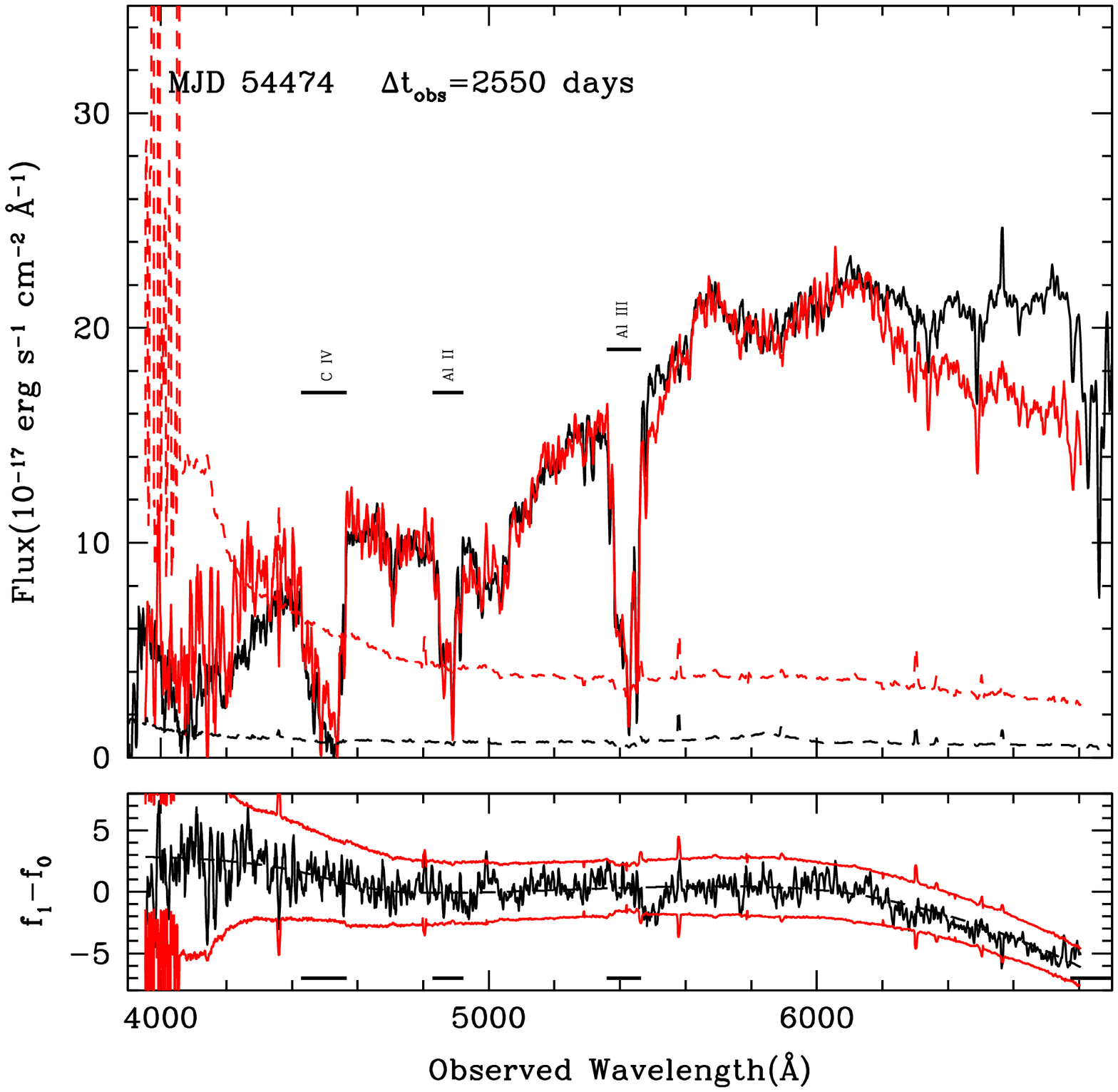,width=9cm,height=6cm}&%,bbllx=21bp,bblly=200bp,bburx=570bp,bbury=705bp,clip=yes}&
\psfig{figure=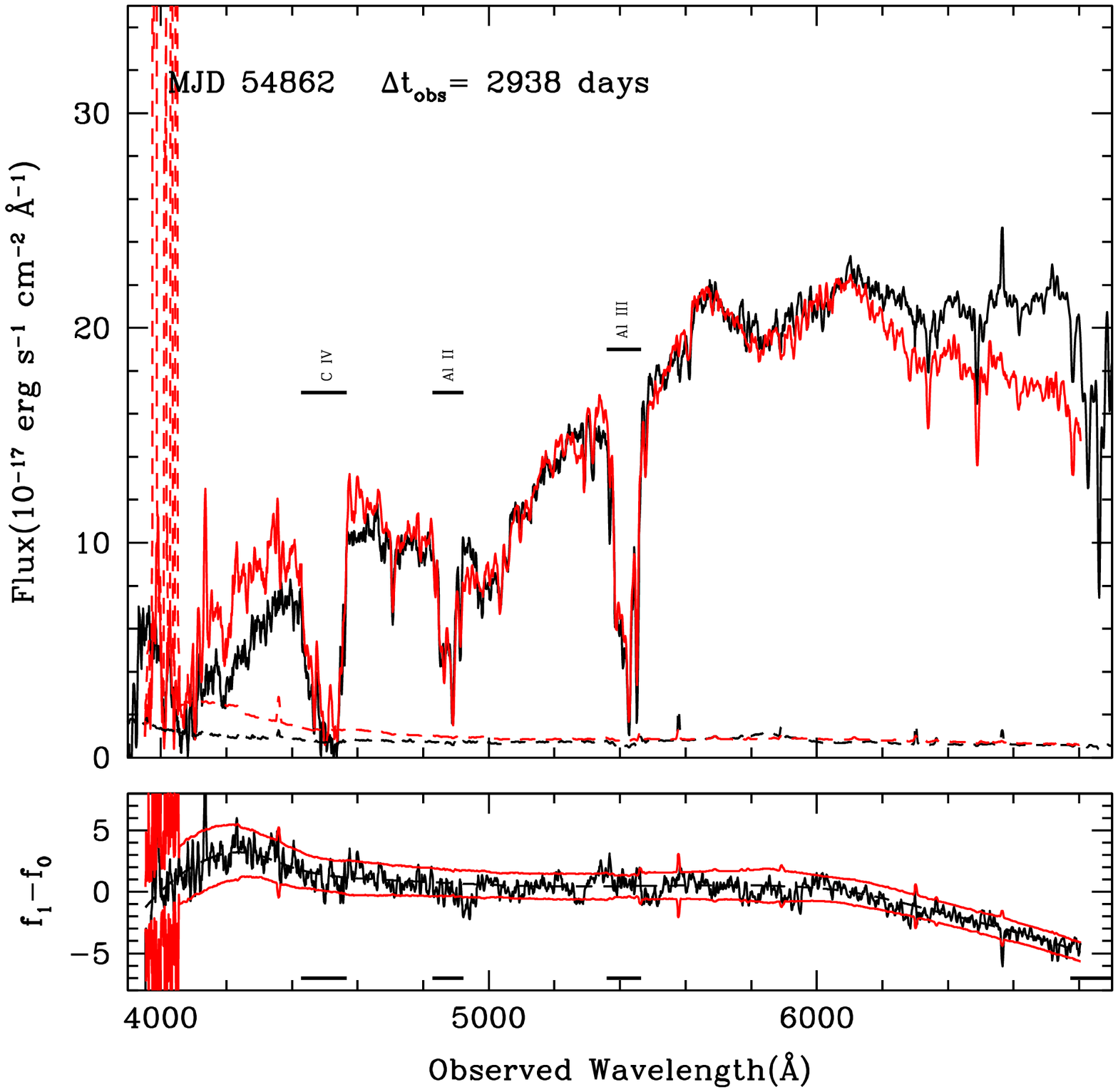,width=9cm,height=6cm}\\%,bbllx=51bp,bblly=200bp,bburx=570bp,bbury=705bp,clip=yes}\\
\psfig{figure=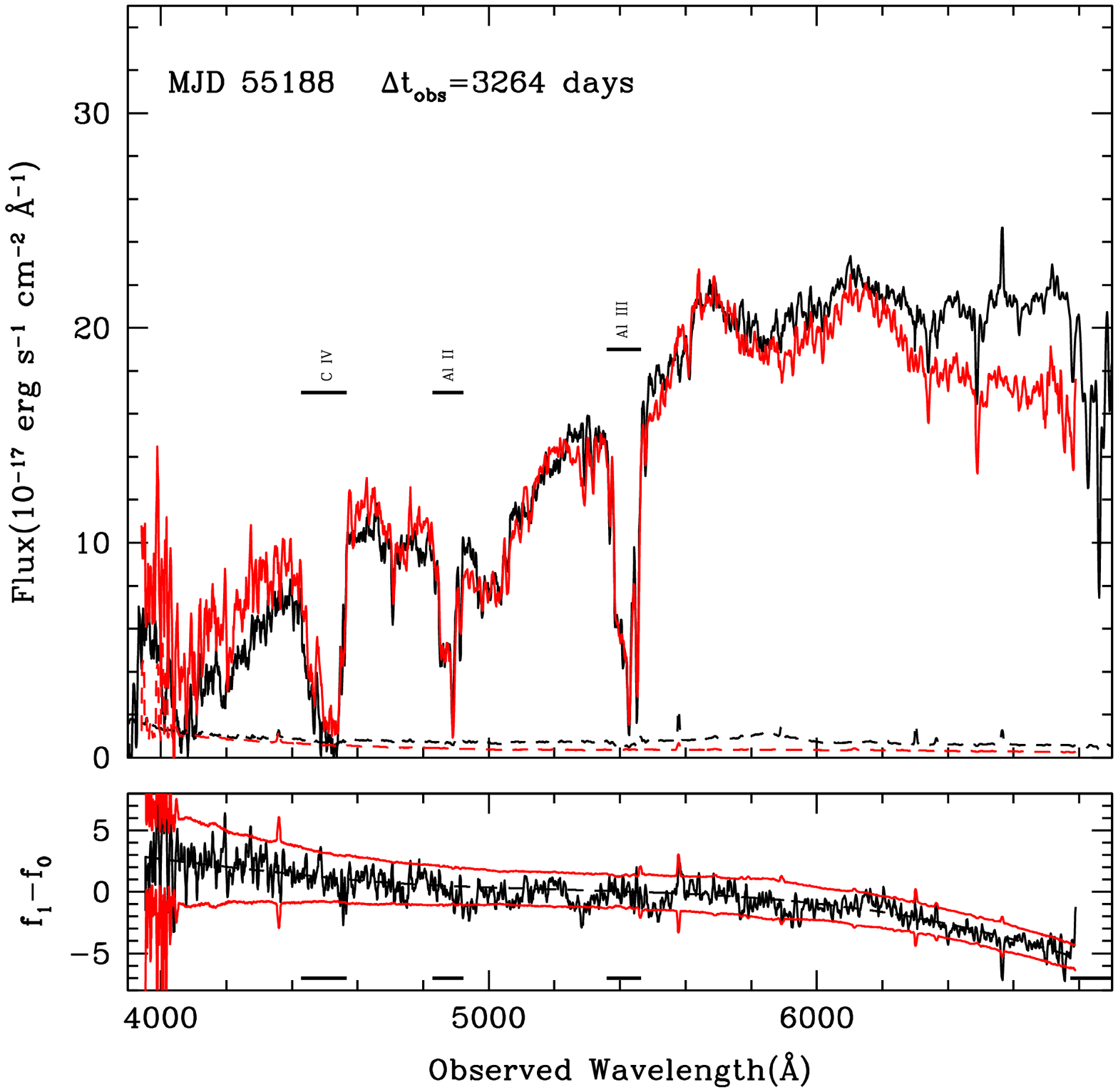,width=9cm,height=6cm}&%,bbllx=21bp,bblly=150bp,bburx=570bp,bbury=698bp,clip=yes}&
\psfig{figure=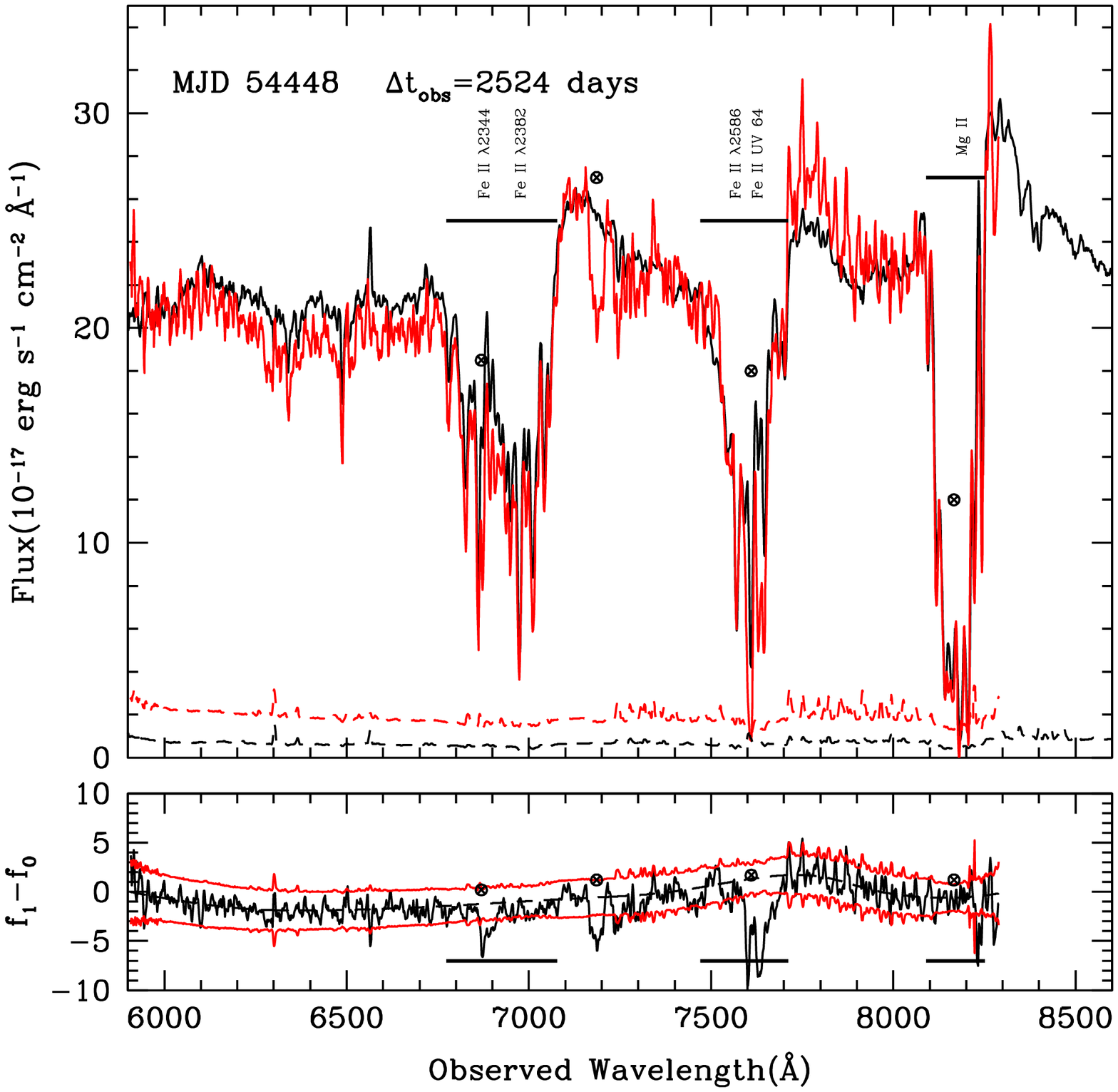,width=9cm,height=6cm}\\%,bbllx=21bp,bblly=200bp,bburx=570bp,bbury=705bp,clip=yes}&
\psfig{figure=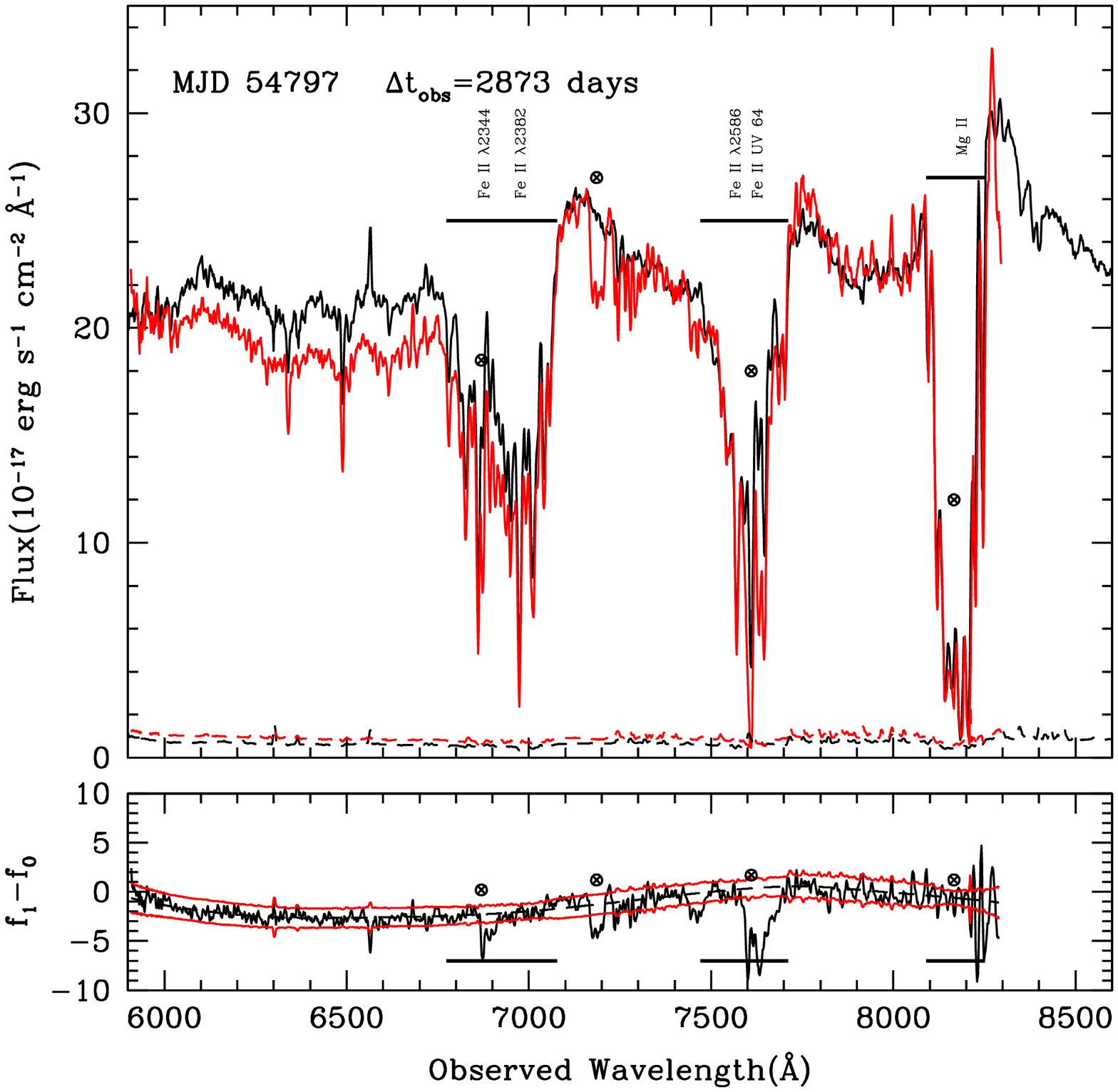,width=9cm,height=6cm}&%,bbllx=51bp,bblly=200bp,bburx=570bp,bbury=705bp,clip=yes}\\
\psfig{figure=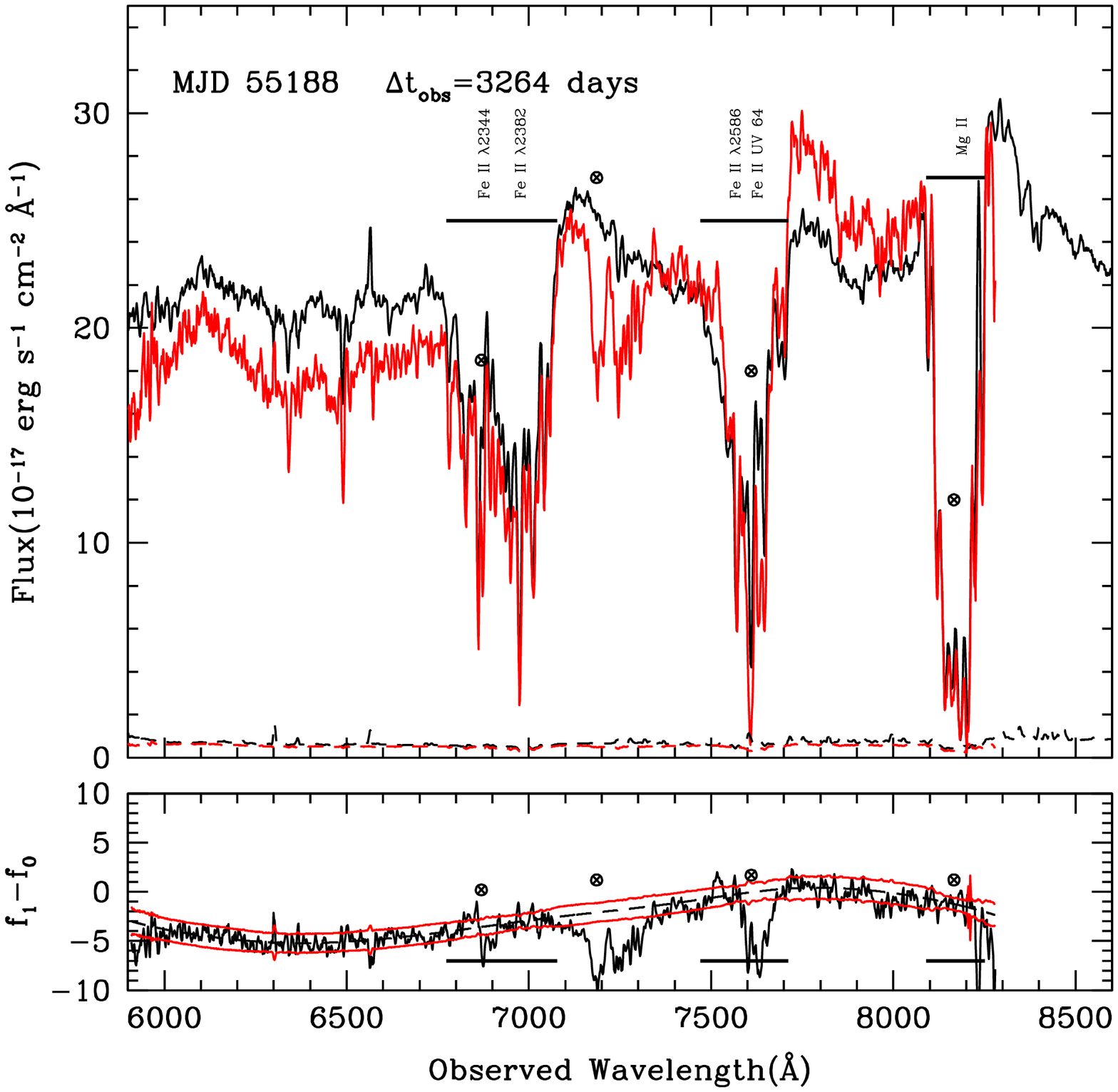,width=9cm,height=6cm}\\%,bbllx=21bp,bblly=150bp,bburx=570bp,bbury=698bp,clip=yes}&

\end{tabular}
\caption{ Spectra of SDSS J0318-0600 from  observed at different epochs at
IGO (plotted in red/gray) are overplotted with the reference SDSS spectra 
(black). The flux scale applies to the reference SDSS spectrum  and all 
other spectra are scaled to match the reference spectrum in continuum. 
%
% In each plot the error spectra are also shown.  The difference spectra for the corresponding MJD's are plotted in the lower panel of each plot. 1 $\sigma$ error is plotted above and below the mean. The regions of absorption lines are marked in the upper panel and the corresponding regions are marked in the lower panel with horizontal lines. 
%
In each plot the error spectra are also shown.  The difference spectra 
for the corresponding epoch together with the associated errors
are plotted in the lower panel of each plot. 
 The regions of absorption lines  are marked  with horizontal 
lines. The regions of strong telluric absorption are marked by crossed circles
.
} 
\label{fig2}
\end{figure*}

%+++++++++++++++++++++++++++++++++++++++++++++++++++++++++++++++++++++++++++++++++++++++++++++++++++++++++++++++
\begin{figure*}
 \centering
\begin{tabular}{c c}
\psfig{figure=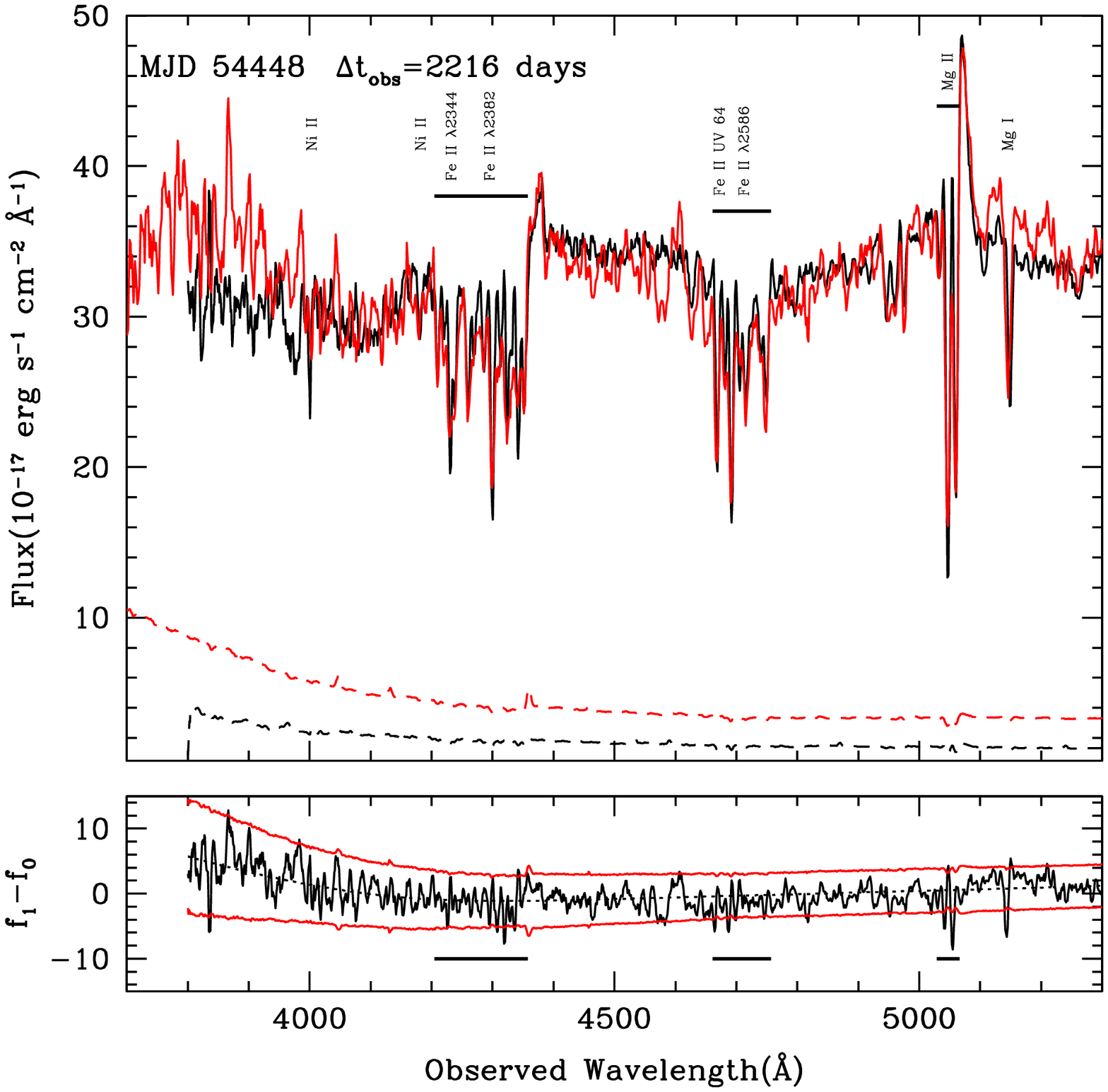,width=9cm,height=6cm}&%,bbllx=21bp,bblly=200bp,bburx=570bp,bbury=705bp,clip=yes}&
\psfig{figure=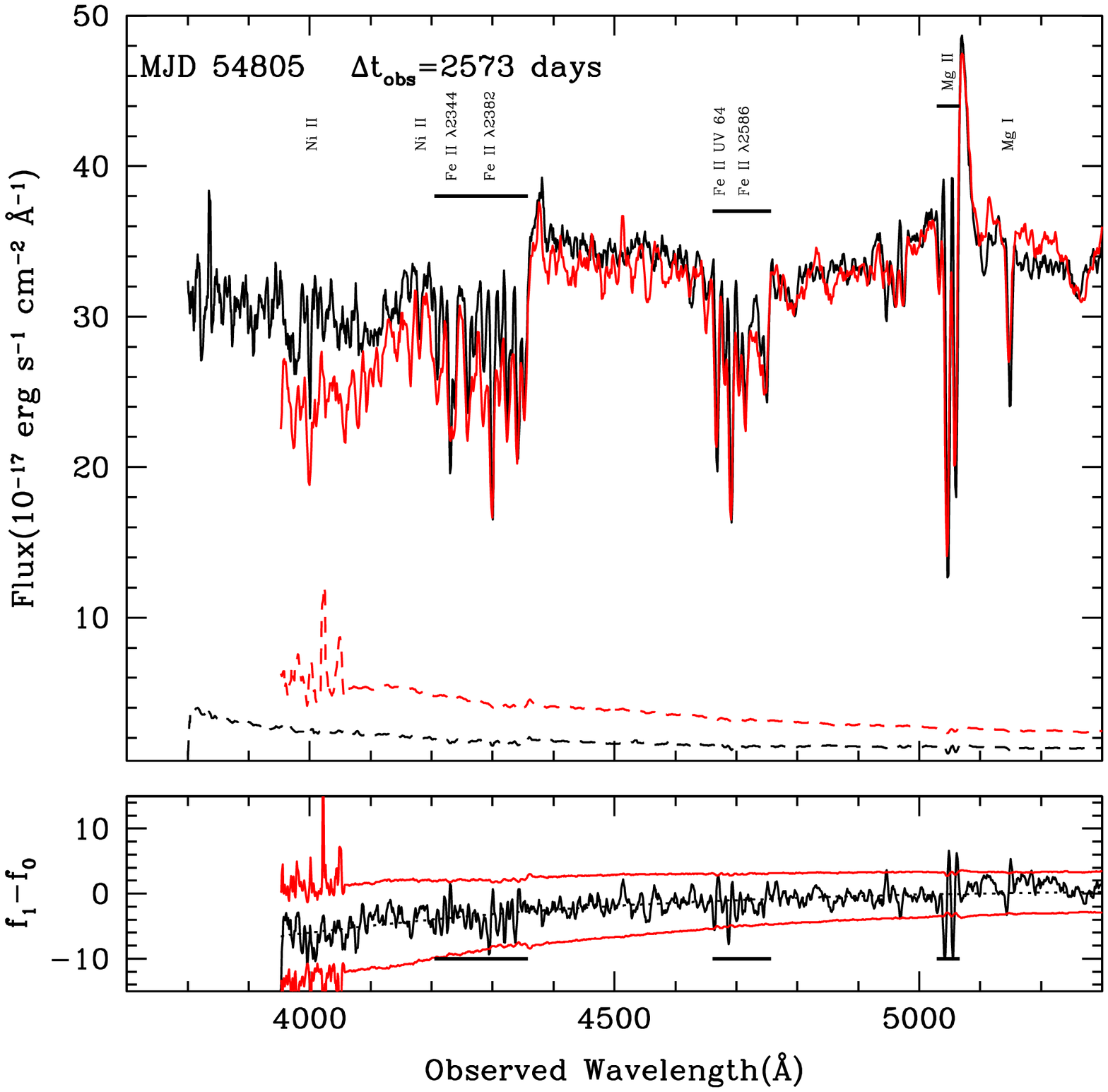,width=9cm,height=6cm}\\%,bbllx=51bp,bblly=200bp,bburx=570bp,bbury=705bp,clip=yes}\\
\psfig{figure=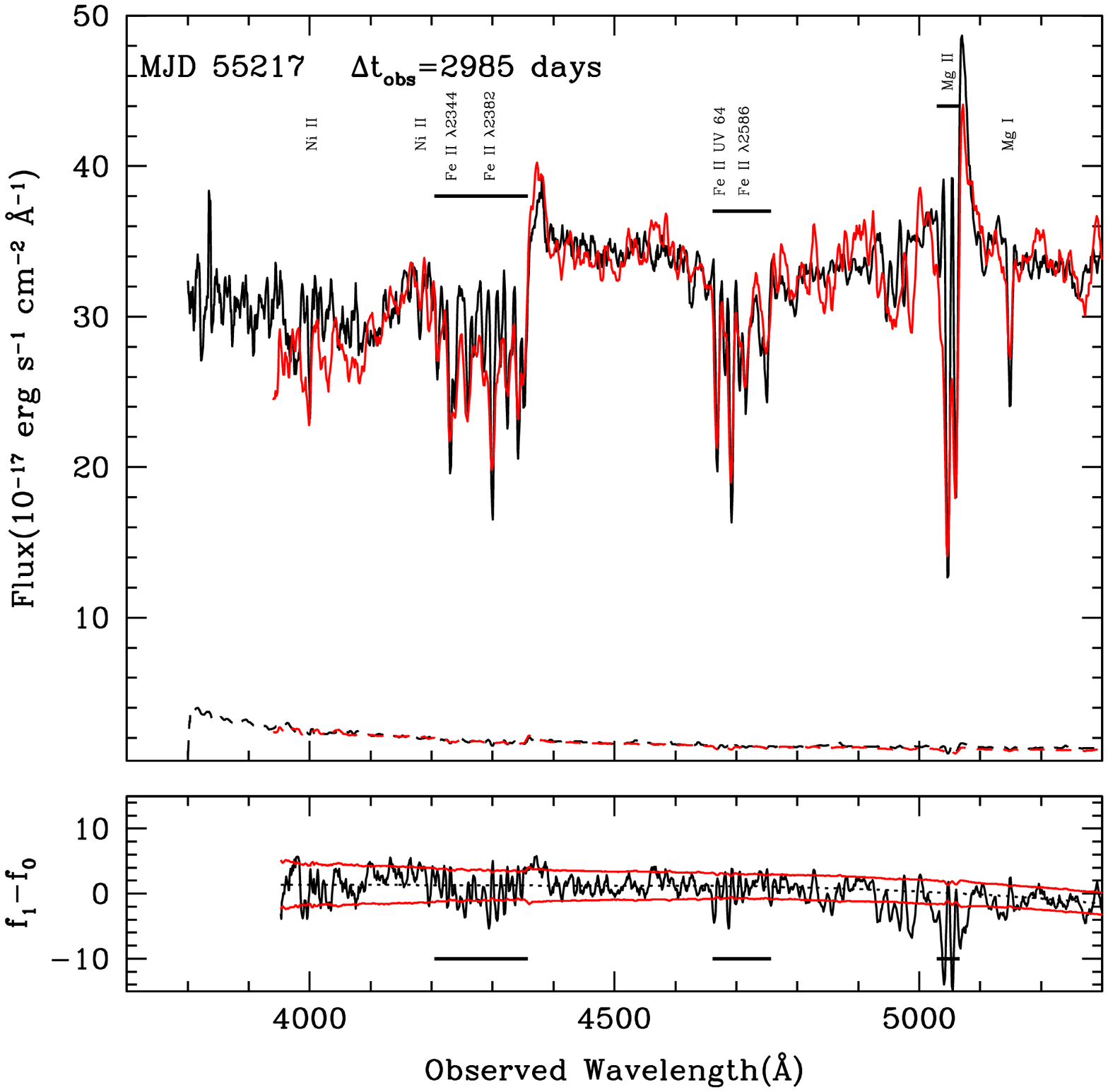,width=9cm,height=6cm}&\\%,bbllx=21bp,bblly=150bp,bburx=570bp,bbury=698bp,clip=yes}&
\end{tabular}
\caption{ Spectra of SDSS J0835+4242  from  IGO (red) epochs are overplotted with the reference SDSS spectra (black). The flux scale applies to the reference SDSS spectrum  and all other spectra are scaled to match the reference spectrum in continuum. 
In each plot the error spectra are also shown.  The difference spectra 
for the corresponding epoch together with the associated errors
are plotted in the lower panel of each plot. 
 The regions of absorption lines  are marked  with horizontal 
lines. The regions of strong telluric absorption are marked by crossed circles.
}
\label{fig3}
\end{figure*}

%+++++++++++++++++++++++++++++++++++++++++++++++++++++++++++++++++++++++++++++++++++++++++++++++++++++++++++++++\begin{figure*}
\begin{figure*}
 \centering
\begin{tabular}{c c}
\psfig{figure=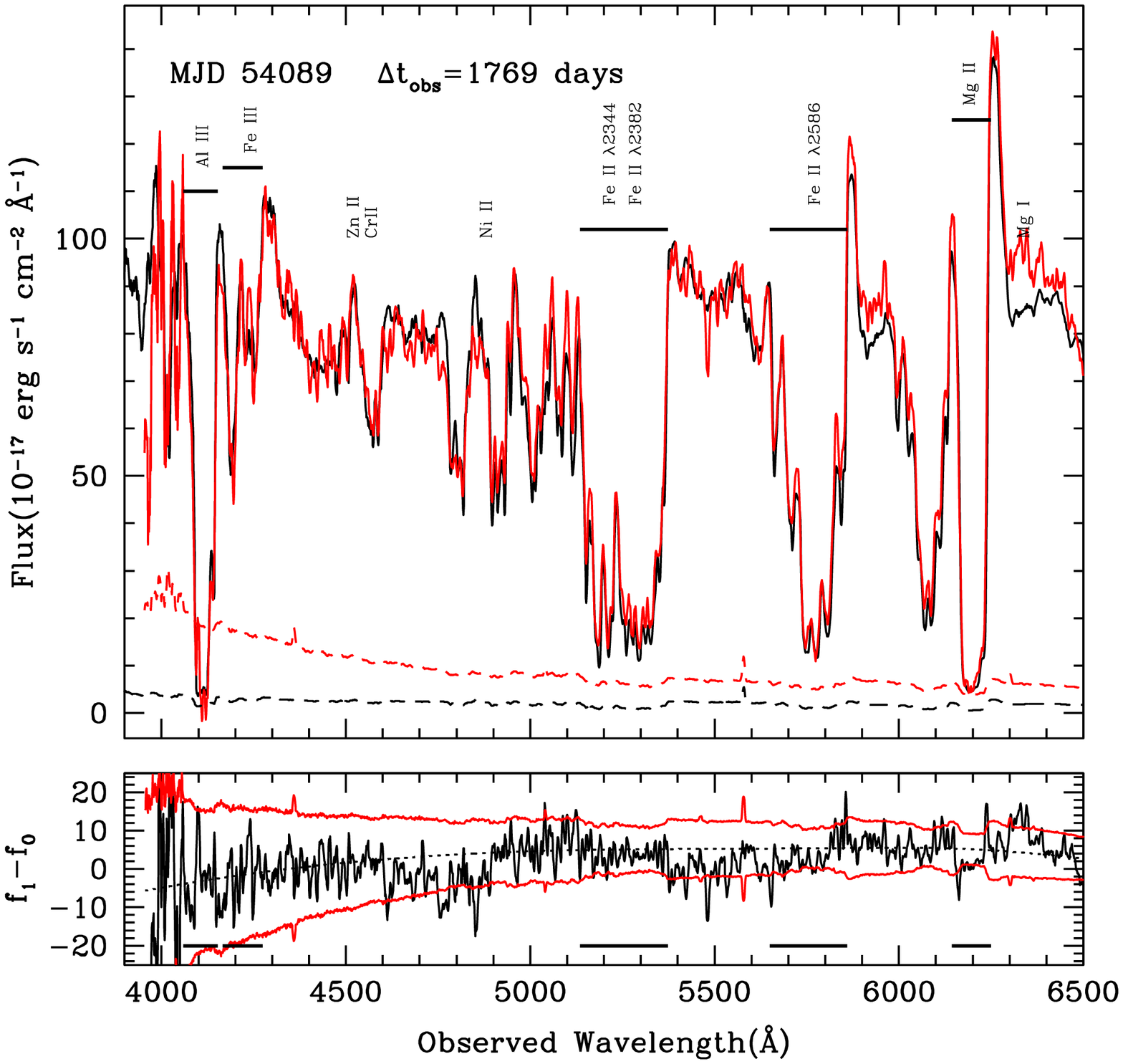,width=9cm,height=6cm}&%,bbllx=21bp,bblly=200bp,bburx=570bp,bbury=705bp,clip=yes}&
\psfig{figure=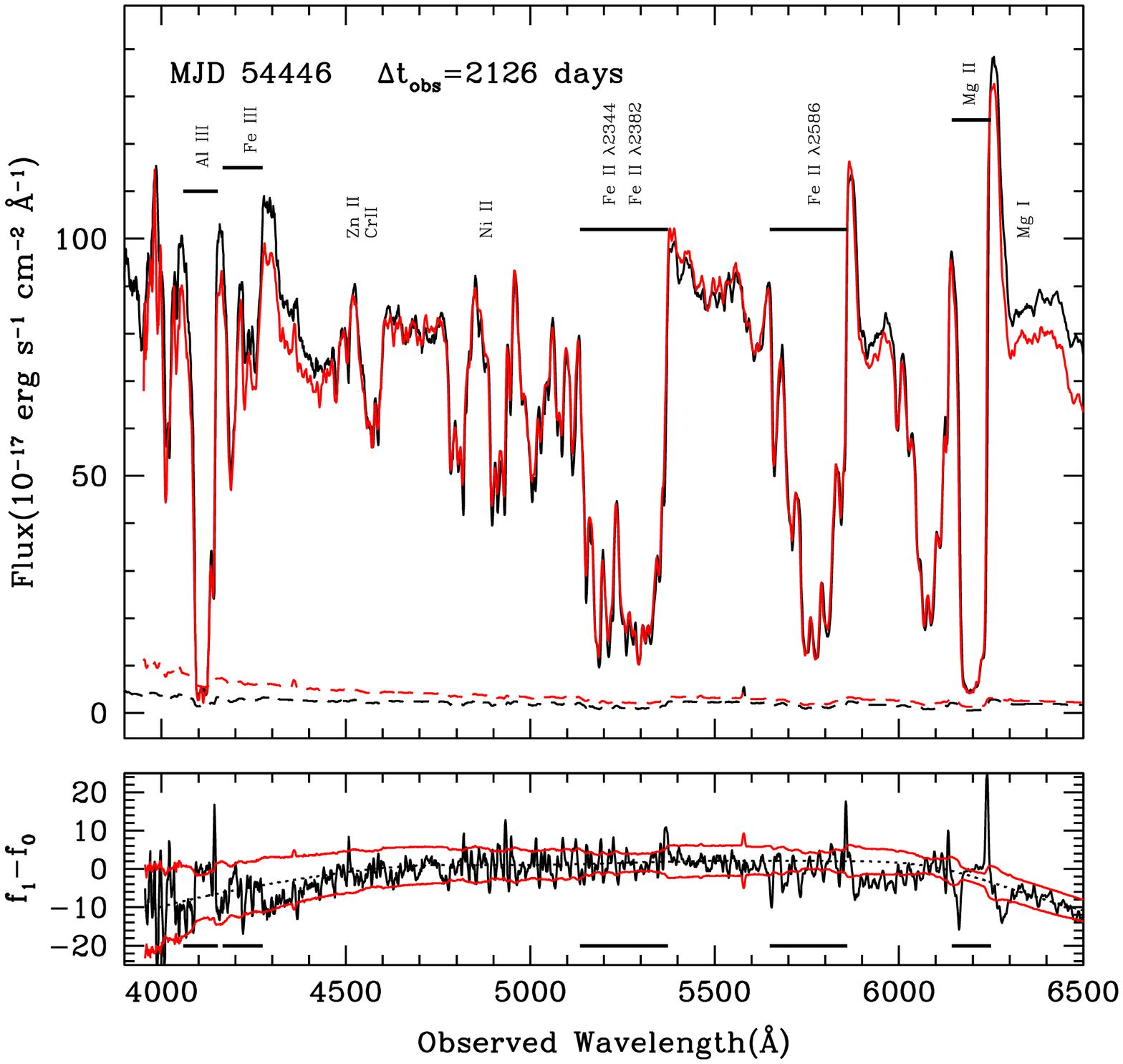,width=9cm,height=6cm}\\%,bbllx=51bp,bblly=200bp,bburx=570bp,bbury=705bp,clip=yes}\\
\psfig{figure=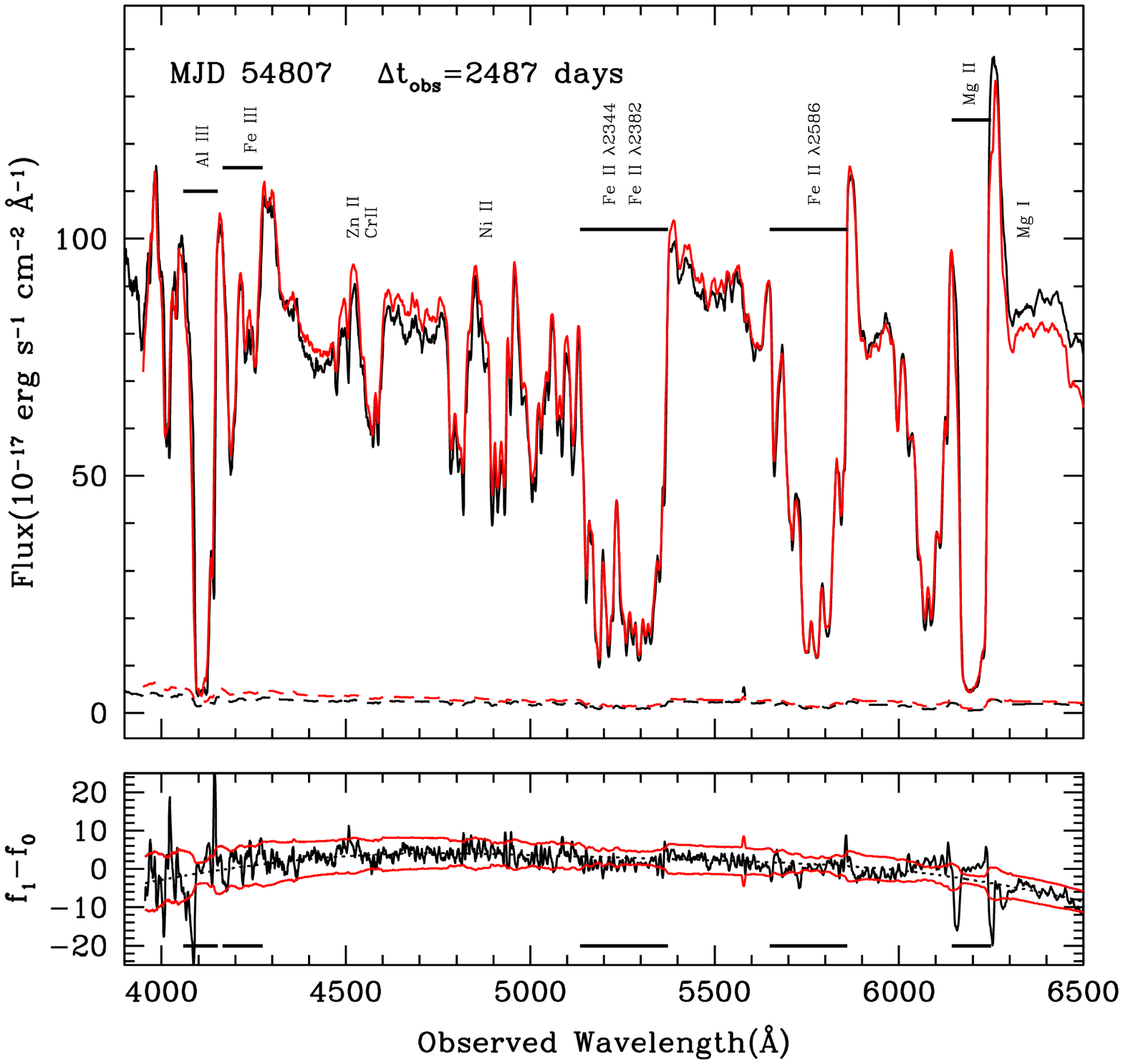,width=9cm,height=6cm}&%,bbllx=21bp,bblly=150bp,bburx=570bp,bbury=698bp,clip=yes}&
\psfig{figure=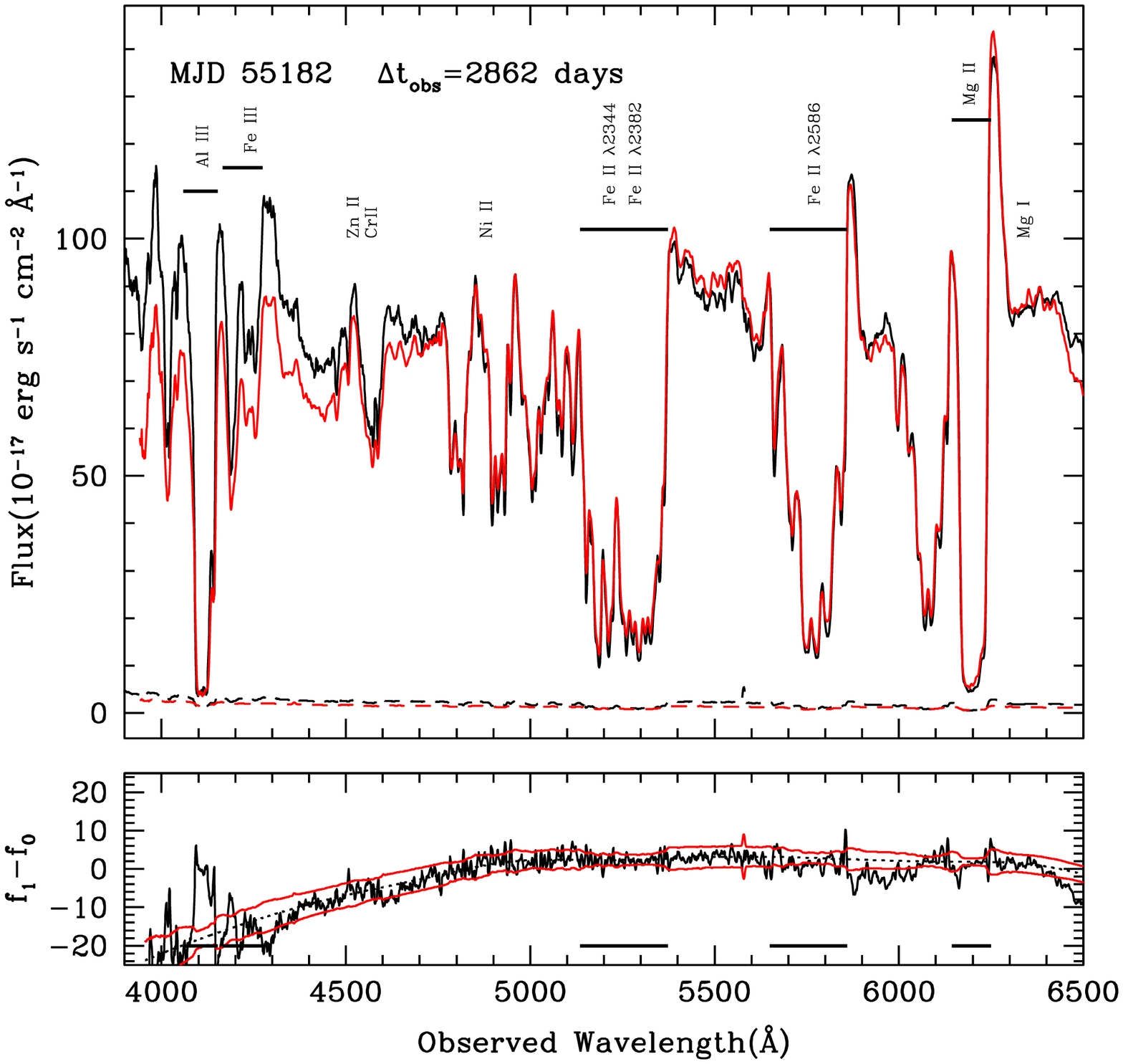,width=9cm,height=6cm}\\%,bbllx=51bp,bblly=200bp,bburx=570bp,bbury=705bp,clip=yes}\\

\end{tabular}
\caption{ Spectra of SDSS J0840+3633  from  IGO (red) epochs are overplotted with the reference SDSS spectra (black). The flux scale applies to the reference SDSS spectrum  and all other spectra are scaled to match the reference spectrum in continuum.
In each plot the error spectra are also shown.  The difference spectra 
for the corresponding epoch together with the associated errors
are plotted in the lower panel of each plot. 
 The regions of absorption lines  are marked  with horizontal 
lines. The regions of strong telluric absorption are marked by crossed circles.
}
\label{fig33}
\end{figure*}

 In Fig.~\ref{fig1} we compare different SDSS and IGO spectra with the 
reference SDSS spectrum observed on MJD 51877.  The spectra are aligned by a 
simple scaling of the mean flux. The spectral ranges covered by absorption
of different species are marked with horizontal lines in each panel.  
It is clear from the Fig.~\ref{fig1} that there are additional absorption 
at the rest wavelength of $\lambda_{obs} \sim$ 4950 \AA\  
that cannot be accommodated by the \feii\ UV63 absorption. This could be due 
to an additional blue shifted \mgii\ component not identified by 
\citet{hall03a}. The corresponding \feii\ UV2 absorption can explain the 
absorption trough at $\lambda_{obs} \sim$ 
4950 \AA\ seen in Fig.~\ref{fig1}.
In bottom of each panel we show the difference between the two spectra
plotted in the upper part together with the associated errors estimated
from the error spectra. If required, in order to take care of spectral
slope differences, we have fitted a lower order polynomial to the
difference spectra considering regions that are devoid of absorption lines. For all the sources, the spectra are smoothed by 5 pixels for better presentation. However, the unsmoothed spectra are used for obtaining the difference and ratio spectrum.

 The two epoch SDSS data and our IGO spectrum obtained in year 2007 
(MJD 54446) are nearly consistent with the reference SDSS spectrum within
measurement uncertainties. We do not find any significant deviations in
the difference spectra at the locations of strong absorption lines. 
However, it must be remembered that  \mgii\ and \feii\ transitions are 
highly saturated. Therefore, we may not be able to detect small column 
density variations. However, part of the absorption by \feii\ fine-structure 
transition UV 63 is unsaturated. So, if there is any variability we should 
have detected the same.
From Fig.~\ref{fig1} it is clear that even this transition does not
show any detectable variability. 
%Even though some spectral differences 
%are seen in the case of IGO
%data they are not statistically significant due to low S/N of the
%spectrum.
%
%
%It can be clearly seen that, all the three systems are present in the \mgii lin%e whereas system A is not present for the \feii lines.

However, the plot suggests some differences between the SDSS and IGO spectrum
obtained in the year 2008.   This is evident from the fact that 
the feature around $\lambda_{obs} \sim$ 4700 \AA\ is consistently 
stronger in the IGO spectrum. The difference spectrum is found to be consistently above zero over 200 pixels suggesting an excess at $>$ 10 $\sigma$ level.
As discussed before, this part of the spectrum never recovers to the expected continuum level and therefore is subject to some absorption. This could be \feiia\  absorption of the possible blue component discussed above.
  Similarly, we see significant deviation at $\lambda\sim5500$ \AA. The
continuum emission in this region is dominated by \feii\ and \mgii\ emission.
Therefore the variability is not related to absorption line variability.

%We have plotted the \citet{vestergaard01} \feii\ emission template in 
%the top panel. It is clear that this regions of spectra have strong \feii\ 
%emissions and no strong absorption line is also expected. Therefore 
%this apparent variability has to be attributed to the changes in the
%strength of the \feii\ emission.  If this is the case then the absorption
%flow has to cover a considerable part of this emission line region for us not
%to detect any variability in the bottom of the absorption lines.
 So, we conclude that major part of BAL absorption lines of 
SDSS J0300+0048 that contains \feii\  fine structure and \caii\  \citep[as identified by][]{hall03} has not varied significantly over a time period 
of $\sim$ 4.2 years in the quasar rest frame. From Fig.~\ref{fig5}, we notice that on an average the V magnitude of this QSO remained same with in 0.05 mag over several years. Thus the photoionization induced variability is not expected as well. However, we report the possible variation of the bluest component.  Future monitoring observations are needed to confirm the nature of the noted variability.

\subsection{SDSS J0318-0600}
SDSS J0318-0600 is a bright, reddened FeLoBAL of emission redshift 
\zem = 1.9669. \citet{bautista10} and \citet{dunn10}  studied the high 
resolution VLT spectrum of this source and identified 11  absorption 
components spanning a velocity range of  $-$7400 to $-$2800 \kms with 
the absorbing clouds fully covering the emitting regions of the background 
QSO. From photoionization models, they report super-solar abundances, 
an electron density of 10$^{3.3\pm0.2}$ cm$^{-3}$ and 
the distance from the emission source to the strongest component as 
6$\pm$3 kpc. The ratio of measured kinetic to bolometric luminosities is 
large enough to consider  this outflow a significant contributor to quasar 
feed back mechanism \citep{faucher11}. The  observed IGO spectra at 
various epochs are overplotted with the reference SDSS 
spectrum (MJD 51924) in  Fig.~(\ref{fig2}). The BAL trough has multiple 
narrow components covering a large wavelength range. We identify four 
distinct components at \zabs = 1.895, 1.911,  1.927 and 1.941 with the 
strongest component being at \zabs = 1.927. The spectrum has well defined 
absorption structures of \siiv, \civ, \alii, \aliii, \mgii, \feii\ 
and its excited states. 

Individual components of \civ\ and \mgii\ lines are saturated \citep[see Figure 3 of][]{dunn10}. As
the line widths are close to our spectral resolution, we smoothed the
SDSS spectrum to IFOSC resolution for comparison. Some of the
variations seen in the case of Fe~{\sc ii}$\lambda$2600 and 
Fe~{\sc ii}$\lambda$2383 are mainly due to atmospheric absorption that
were not corrected in the IFOSC spectra. These regions of strong telluric 
absorption are marked in the spectra by crossed circles. 

 The prominent absorption components are labeled in Fig.~\ref{fig2} and
the extent of the absorption are shown with horizontal lines. We notice
continuum shape differences between IGO and SDSS spectra. To take 
care of this we fit a smooth lower order polynomial to the difference
spectrum avoiding the absorption line regions and regions affected by
telluric lines.
It is clear from the bottom part of each panel that there is no 
significant deviation spread over the wavelength range covered by 
absorption in any of the cases. Strong deviations are seen only 
in regions where there is telluric contaminations in our IFOSC spectrum.
  
% Given the telluric contamination in the  
%Fe~{\sc ii}$\lambda$2600 region and saturation at \civ, no conclusive 
%evidence for BAL variations can be inferred for this source. 
Hence, we 
conclude that SDSS J0318-0600 shows no significant variability over a 
time scale of ~ 3 years in the quasar rest frame. The light curve presented 
in Fig.~\ref{fig5} suggests a smooth dimming of the QSO (by $\lesssim$ 0.1 mag). 
Our IGO spectra have slightly low continuum flux compared to SDSS spectrum
in the wavelength range 6000$-$7000 \AA. This could be related either to
the uncertainty related to extinction corrections or spectral variability
of the QSO. However, this does not affect our conclusion regarding the
absence of variability in the absorption lines.

\subsection{SDSS J0835+4242}

 \begin{table}
\caption{Equivalent width measurements  for absorption lines
seen towards SDSS J0835+4242:}
\begin{tabular}{ccccc}
  \hline
  &\multicolumn{4}{c}{Equivalent widths}\\
Spectrum&\multicolumn{4}{c}{(\AA)}\\
  &Mg II&Mg I &Fe II&Fe II\\
&&	&$\lambda$2586\AA&$\lambda$2382\AA \\
  \hline
  \hline
  SDSS&       4.1$\pm$0.2&	0.7$\pm$0.2     &    5.2$\pm$0.3    &      10.7$\pm$0.3 \\ 
 IGO-2007&    4.5$\pm$0.3&	0.6$\pm$0.2     &    5.5$\pm$0.7    &      11.5$\pm$1.0\\ 
 IGO-2008&    4.5$\pm$0.3&	0.6$\pm$0.2     &    6.2$\pm$0.9    &      11.4$\pm$1.4 \\ 
  IGO-2010&   5.6$\pm$0.2&	0.5$\pm$0.2     &    4.5$\pm$0.2    &      11.2$\pm$0.9 \\
 
  \hline
  \end{tabular}
\label{tab0835}
% \caption{Equivalent width measurements with one sigma error for absorption
%seen towards SDSS J0835+4242:}
  \end{table}

The redshift of this BAL QSO is \zem = 0.810.
Fig.~\ref{fig3} shows the comparison between the reference SDSS and IGO data for SDSS J0835+4242.
IGO data acquired in the year 2010 (MJD 55217) has slightly lower resolution compared
to the data taken in 2008 (MJD 54448). The differences between IGO and SDSS spectra seen in the top panel of Fig.~\ref{fig3} are mostly a consequence of this. 
The spectrum contains strong absorption lines from \mgii, \mgi, \mnii, \ni2\
and  \feii\ multiplet lines like Fe~{\sc ii}$\lambda 2586$, Fe~{\sc ii}$\lambda 2382$, Fe~{\sc ii}$\lambda 2344$ and excited UV 64. 
\mnii$\lambda 2576$ is blended with the \feii$\lambda 2586$. 
The quasar has strong absorption troughs that reach peak depth at  
\zabs = 0.805.  Like the previous system the absorption trough
is resolved into multiple narrow components. Additional \mgii\ 
absorption systems are also seen at \zabs = 0.800 and \zabs = 0.769. 
The associated ground state \feii\ lines for these two systems are not detected clearly in our spectrum.
\
 As absorption lines are narrow, the continuum
can be well defined. So, we measured the total equivalent widths of 
\mgii, \mgi\ and the blends near Fe~{\sc ii}$\lambda$2586 and  Fe~{\sc ii}$\lambda$2382. The results are summarized in Table~\ref{tab0835}. The quoted error in the equivalent widths also
includes uncertainties associated with the continuum placement calculated using repeated 
continuum fits. This table also confirms the lack of significant (ie. $>$ 5 $\sigma$ level) absorption line variability. 

We conclude that absorption in SDSS J0835+4242 has not varied significantly over 
the time period of 4.97 years in the quasar rest frame. The light curve presented in Fig.~\ref{fig5} also suggests that on an average the QSO has not varied in its brightness by more than 0.1 mag.

\subsection{SDSS J0840+3633}
 SDSS J0840+3633 is one among the  first known radio loud BAL QSOs \citep{becker97}. In the discovery paper, \citet{becker97} pointed out the correlation between the column densities of low ionization clouds and radio emissions for 3 LoBALs and went on to suggest that LoBALs may be transition objects between radio loud and radio quiet BAL QSOs.  \citet{brotherton97} performed the spectropolarimetry of this source and found it to be a highly polarized BALQSO where the continuum polarization rises steeply toward shorter wavelengths while keeping a constant position angle in the continuum. They consider scattering as the likely polarization mechanism, with the effects of some combination of dust and dilution leading to the wavelength dependences seen. Their studies showed that SDSS J0840+3633 has unpolarized emission lines and increased polarizations in its BAL troughs but complex polarization behavior across its narrow metastable troughs.
\citet{dekool02} studied the KECK/HIRES spectrum of this quasar. Their spectrum reveals outflowing gas with two main components. The physical conditions in the two components are found to be significantly different. This is attributed to the difference of a factor of $\sim$ 100 in the distance from the central source. The low velocity gas has absorption from excited states which indicate low density. Assuming UV fluorescence as the possible excitation mechanism, they estimate the distance between the low velocity absorber and the active nucleus to be $\sim$230 pc. The high velocity high density gas gives rise to strong \feiii\ and \aliii\ lines.  The estimated distance between this gas and the nucleus is $\sim$ 1  pc.

	Each panel of Fig.~\ref{fig33} shows the IGO spectrum obtained at different epochs  overplotted  with the reference SDSS spectrum. The spectrum is completely dominated by absorption lines from Mg~{\sc ii}, Ni~{\sc ii}, Cr~{\sc ii}, Al~{\sc iii}, Si~{\sc ii}, Al~{\sc ii} and excited states of \feii\ and Fe~{\sc iii}. The unabsorbed \mgii\ emission gives the emission redshift to be 1.230. The BAL troughs are nearly saturated and the continuum is heavily absorbed. The low velocity \mgii\ absorption component has a redshift of \zabs = 1.225 and the troughs extends to a velocity of about 4000 km s$^{-1}$.  Excited states of \feii\ comprises of UV63 lines and that of \feiii\ comprises of UV34 lines. 

 From the difference spectrum plotted in the lower half of each panel, we see no consistent variations in  \mgii, \feii,  or \feiii\ absorption lines. However, the difference spectra have a smooth curvature suggesting a possible differences in the continuum. So, we conclude that the BALs in SDSS J0840+3633 has not varied by an appreciable amount over a rest frame time period of $\sim$ 4 years. From the light curve presented in Fig.~\ref{fig5}, we notice a gradual increase in the V-magnitude of the QSO. But the changes are within 0.05 mag. This, together with  the strong saturation of the BAL troughs, could be the reason for the lack of line variability.
 \begin{figure*}
\centering
\begin{tabular}{c}
\psfig{figure=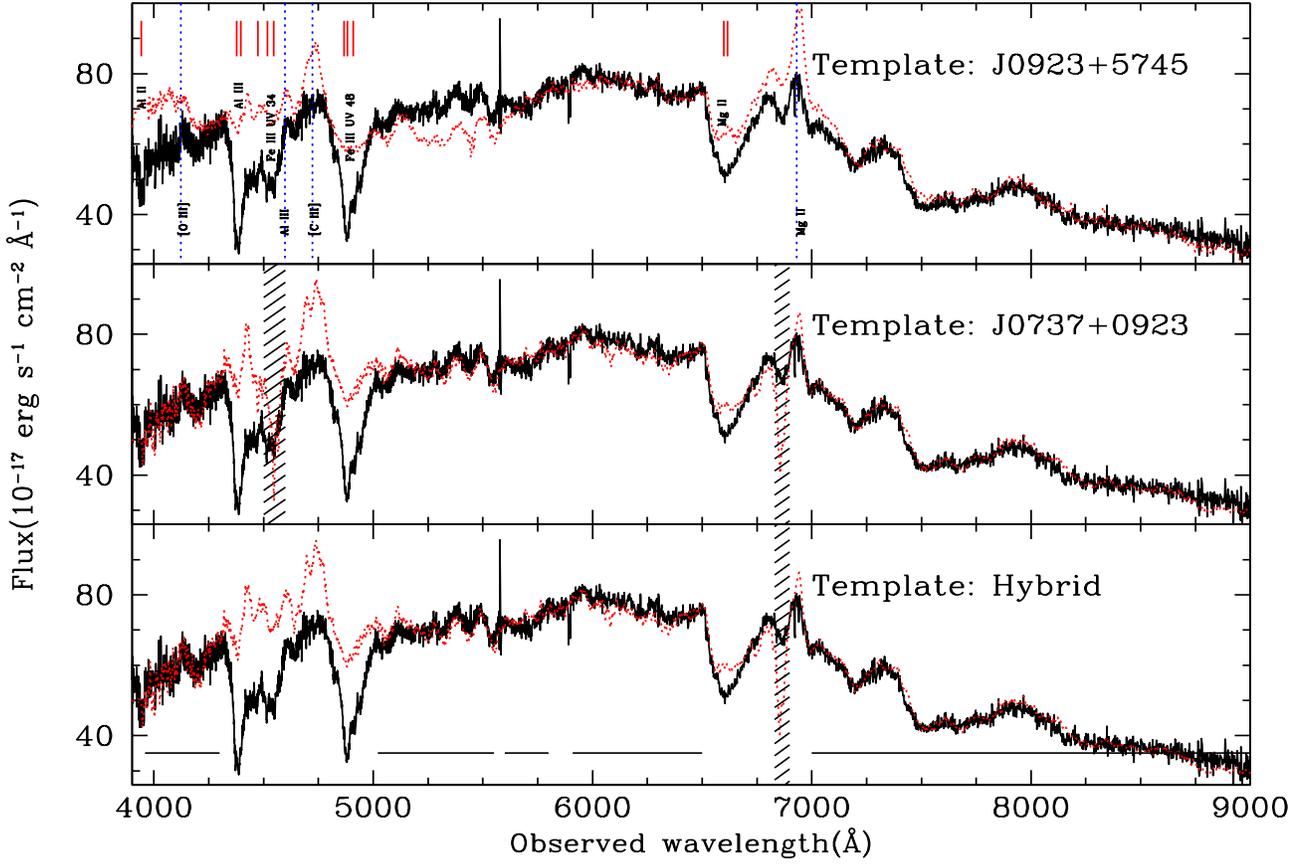,width=1.0\linewidth,height=0.7\linewidth,angle=270}
\end{tabular}
\caption{Fitting the spectral energy distribution of SDSS J 2215-0045
observed on MJD 51804.
The prominent absorption and emission lines are identified with 
ticks and dotted lines respectively.  The wavelength bins used for the fitting are shown as horizontal lines in the bottom panel. The top and middle panels show
the best fitted continuum using the spectra of J0923+5745 and J0737+0923
as templates respectively. The SED is well fitted when we use the spectrum of  J0737+0923
as the template. The shaded regions in the middle and bottom panels show
the wavelength range affected by absorption in the spectrum of J0737+0923.
We use the spectrum of J0923+5745 to reconstruct the SED in these narrow
wavelength range. Fit to the spectrum of SDSS J2215-0045 using this
hybrid template is shown in the bottom panel. } 
\label{fig44}
\end{figure*}

 \begin{figure*}
\centering
\begin{tabular}{c}
\psfig{figure=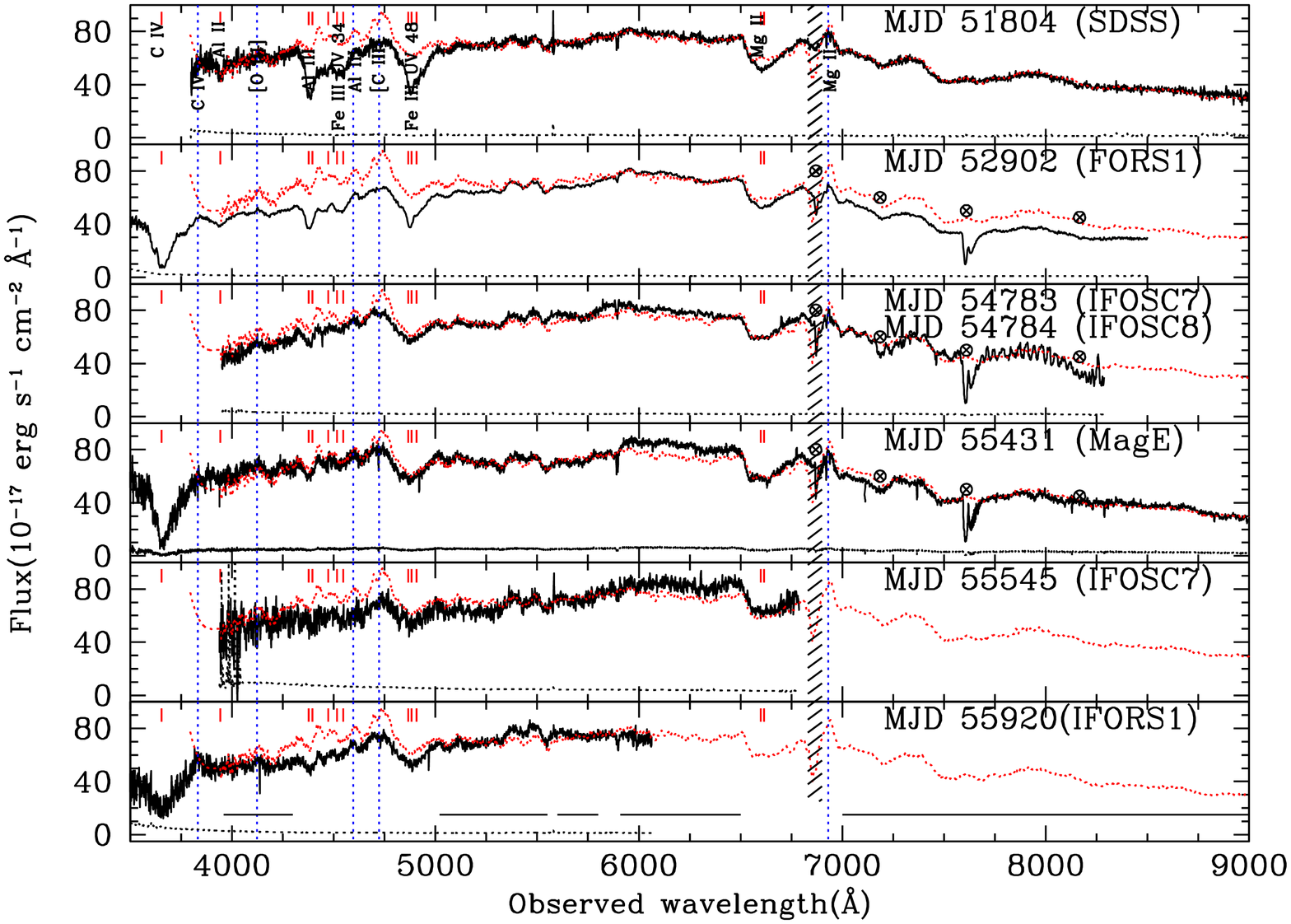,width=1.0\linewidth,height=0.7\linewidth,angle=270}
\end{tabular}
\caption{Spectra obtained for SDSS J2215-0045 at different epochs (together with 
the error spectra) are overplotted with the fit obtained for SDSS  using the hybrid template.
The prominent absorption and emission lines are identified with 
ticks and dotted lines respectively.  The wavelength bins used for the fitting are shown as horizontal lines in the bottom panel. Regions of telluric absorption are marked with crossed circles. The shaded regions in  all the panels show
the wavelength range affected by absorption in the spectrum of J0737+0923.} 
\label{J2215variable}
\end{figure*}

\subsection{SDSS J2215-0045}

This is one of the peculiar QSOs identified by \citet{hall02} 
with strong  \feiii\ UV48$\lambda${2080}, UV34$\lambda${1910} and  
Al~{\sc iii} absorption lines in addition to the strong C~{\sc iv} 
broad absorption line at \zabs$\sim$ 1.36. Fe~{\sc ii} lines are absent 
and the QSO continuum seems to be dominated by the broad emission from
Fe~{\sc ii} and Fe~{\sc iii}. This is supported by the fact that the QSO 
continuum does not have strong intrinsic polarization \citep{dipompeo11}. 
The associated Mg~{\sc ii} absorption coincides with the 2670 \AA\ break 
of the Fe emission template thus its strength can not be accurately 
measured.
The three unusual aspects of this QSO are (i) the lack of \feii\ absorption
while absorption from \feiii\ fine-structure levels are clearly detected 
(ii) the 
\feiii\ UV48 (EP=5.08 eV) absorption line being stronger 
than the \feiii\ UV34 (EP=3.73 eV) which is unphysical under LTE assumption  
and (iii) the emission lines are weak and the redshift of the QSO
is determined by a weak and narrow Mg~{\sc ii} line at \zem = 1.478.

In addition to the broad absorption lines discussed above, a
 narrow associated absorption line system is detected at redshift 
\zabs = 1.4752 based on the presence of Mg~{\sc ii}$\lambda\lambda$2796,2803, 
Fe~{\sc ii}$\lambda\lambda\lambda$2600,2586,2382, 
Si~{\sc ii}$\lambda$1526, C~{\sc ii}$\lambda$1334  in the VLT/UVES data.
Fe~{\sc ii}$\lambda$2344 line falls in the red edge of the 
Na D$_{1}$ and D$_2$ lines from our galaxy. No absorptions 
from excited levels like C~{\sc ii}$^*$ $\lambda$1335 or 
Si~{\sc ii}$^*$ $\lambda$1533 are detected in the spectrum. The VLT/UVES 
spectrum reveals a second \civ\ narrow absorption system at a redshift 
\zabs = 1.07464. The associated \siiv\ is beyond the UVES coverage and  
other associated absorption lines are not detected for this second system. 
We are unable to  confirm these narrow lines in our other datasets due to 
poor spectral resolution. Hence, we are limited to carry out the time 
variability studies for the broad absorption lines only.

Based on photoionization models
%CLOUDY 
%modeling of 
\citet{dekool02} have shown that Fe~{\sc iii} column density
being higher than that of Fe~{\sc ii} can be easily produced in a 
high density outflow (log~$n_{\rm H}$(cm$^{-3}$) $\geq$ 10.5  for 
ionization parameter log U$\simeq-2$) with $N$(H) in a 
narrow range such that the outflow is constituted  only by a 
Fe~{\sc iii}+Al~{\sc iii} zone without having the low ionization 
zone that usually contains Fe~{\sc ii}. \citet{rogerson11} have not 
detected X-ray emission from this source in their Chandra observations. 
They found  the logarithm of the ratio between the 2 keV (l$_{\rm 2~keV}$ ) 
and 2500\AA (l$_{2500\AA}$) rest frame specific luminosities,
$\alpha_{\rm ox}\le -2.45$ and an absorption column of $N$(H) 
$\ge$ 3.4$\times10^{24}$ cm$^{-2}$.  They argued that the optical 
absorption originates from different gas than the one that produces 
X-ray absorption. 
Lyman-$\alpha$ photons can pump  Fe~{\sc iii}$\lambda$1914 
\citep{johansson00}.  This process may explain the weakness of UV34 
absorption. On the contrary due to strong Fe~{\sc ii}~
emission at 1800 \AA\ and the lack of emission at 2000 \AA\, the
observed spectrum may be consistent with both Fe~{\sc iii} UV 48 and
UV 34 being saturated with a covering factor of 0.35 \citep[see][]{hall03}.  
From Fig.~\ref{fig5} we see that this is the source with appreciable 
flux variability in our sample.

While the QSO looks normal in the observed wavelength range 
$\ge$ 6000{\AA} there are clear reddening signatures in the blue. 
As pointed out by  \citet{hall03a} continuum of this QSO 
in the wavelength range between 5000$-$9000{\AA} is dominated by 
Fe~{\sc ii} and Fe~{\sc iii} emission. Fig.~\ref{fig44} shows the 
spectral energy distribution fitting of SDSS J2215-0045 observed 
on MJD 51804. We fit the observed spectrum $f$($\lambda$) with the 
template spectrum $f_{\rm t}$($\lambda$) using the following parametrization,
\begin{equation}
f(\lambda) = \left[ a f_t(\lambda) + b \left( \frac{\lambda}{\lambda_0}\right)^{\alpha} \right] e^{-\tau_{\lambda}}
\end{equation}
We use the $\chi^2$ - minimization to get the best values for parameters  
a, b, $\alpha$ and $\tau_{\lambda}$. The second term in the above equation 
is to take care of the spectral index differences between the observed and 
the template QSO spectra. The dust optical depth `$\tau_{\lambda}$' is 
obtained for the SMC like extinction curve.
The fitting method is similar to the one described in \citet{anand08}.
The wavelength ranges used for the fitting are shown as horizontal 
segments in the top panel. The spectrum has absorption from resonance 
lines like Mg~{\sc ii}, Al~{\sc ii},  \aliii\ and 
fine structure  lines \feiii\ UV 34 and  \feiii\ UV 48. \civ\ and \siiv\ BAL 
are seen in the VLT/UVES spectrum which has a good coverage in the blue. 
The position of absorption lines are indicated by vertical marks in 
Fig.~\ref{fig44} The emission spectrum is illustrated by a dotted line. 
The top and middle panels show the best fitted continuum using the spectra 
of significant iron emitters J0923+5745 and J0737+0923
as templates respectively. The SED is well fitted when we use the spectrum of  J0737+0923
as the template. The shaded regions in the middle and bottom panels show
the wavelength range affected by absorption in the spectrum of J0737+0923.
We use the spectrum of J0923+5745 to reconstruct the SED in these narrow
wavelength range. Fit to the spectrum of SDSS J2215-0045 using this
hybrid template is shown in the bottom panel.  The SED fitting results in a E(B-V) value of -0.083 for the SDSS spectrum. The plot shows that the actual continuum follows the  hybrid template reasonably well.

A weak Al~{\sc ii} absorption is detected at the redshift of 
the \aliii\ and \feiii\ lines. Al~{\sc iii} absorption seems very 
strong. As pointed out by \citet{hall02} the Fe~{\sc iii} UV34 
blend seems to be weaker than UV48, if we use a smooth continuum to 
SDSS J2215-0045. However, it is clear from Fig.~\ref{fig44} that 
apparent weakness of  UV34 may be related to the shape of the QSO 
spectral energy distribution that shows a strong dip at the location 
of Fe~{\sc iii} UV48 absorption. The hybrid template used above suggests 
a stronger [C~{\sc iii}] emission line compared to what is observed for 
SDSS J2215-0045. Removal of this contribution may further reduce the 
continuum level compared to what is observed for SDSS J2215-0045.  
Therefore apparent inconsistencies could be the artifact of the 
unknown continuum shape and need not be related to the population 
inversion by some non-equilibrium process. Fig.~\ref{J2215variable} 
shows the spectra obtained at different epochs overplotted with the 
fit obtained for the SDSS data using the hybrid template.

%++++++++++++++++++++++++++++++++++++++++++++++++++++++++++++++++++++++++++++++++++++++++++++++++++++++++++++++
%+++++++++++++++++++++++++++++++++++++++++++++++++++++++++++++++++++++++++++++++++++++++++++++++++++++++++++++++
\begin{figure*}
 \centering
\begin{tabular}{c}
\psfig{figure=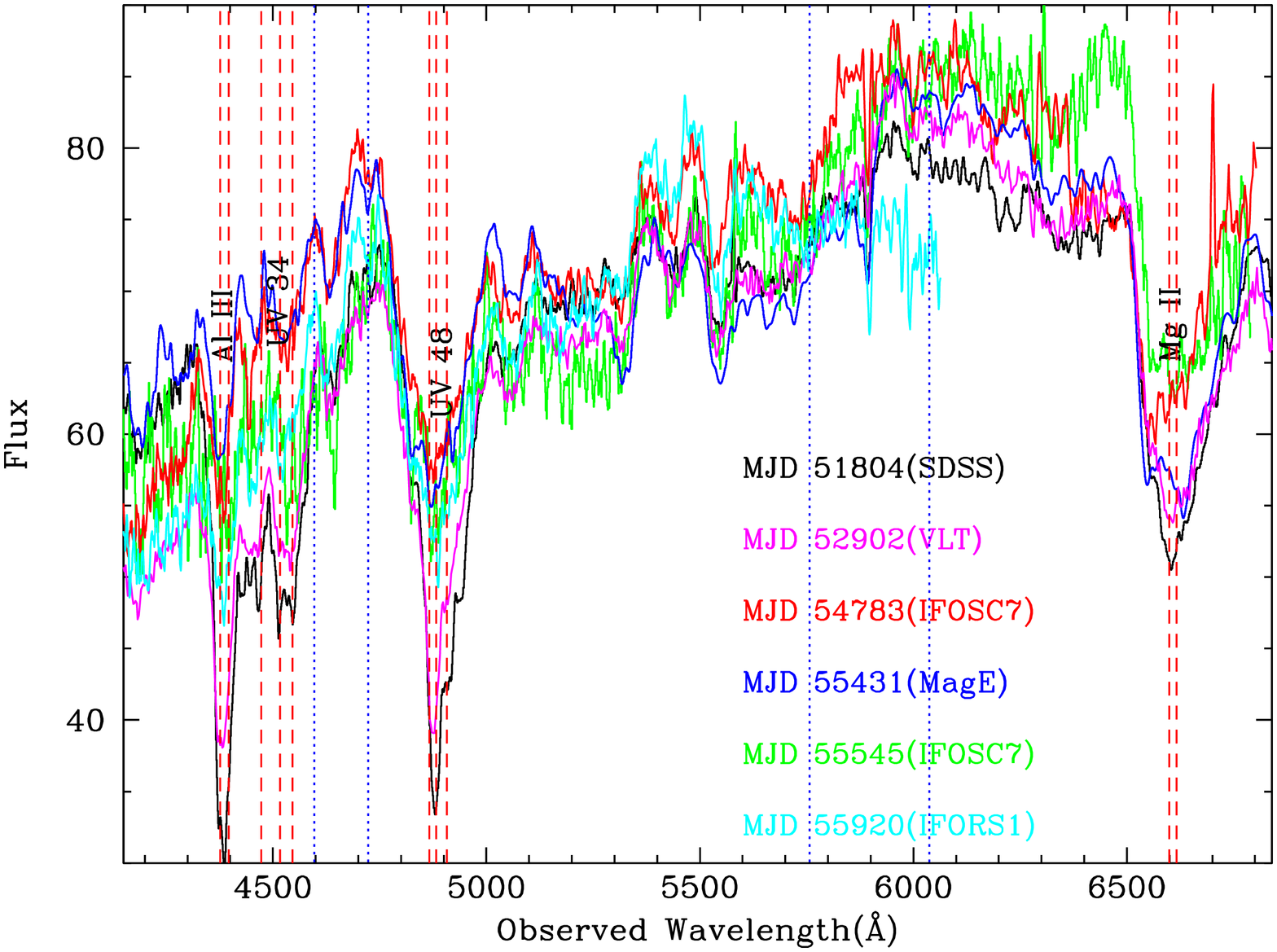,width=1.0\linewidth,height=0.4\linewidth,angle=270}\\%,bbllx=290bp,bblly=0bp,bburx=41bp,bbury=740bp}\\%bbllx=760bp,bblly=30bp,bburx=26bp,bbury=268bp,clip=true}\\ 
 \psfig{figure=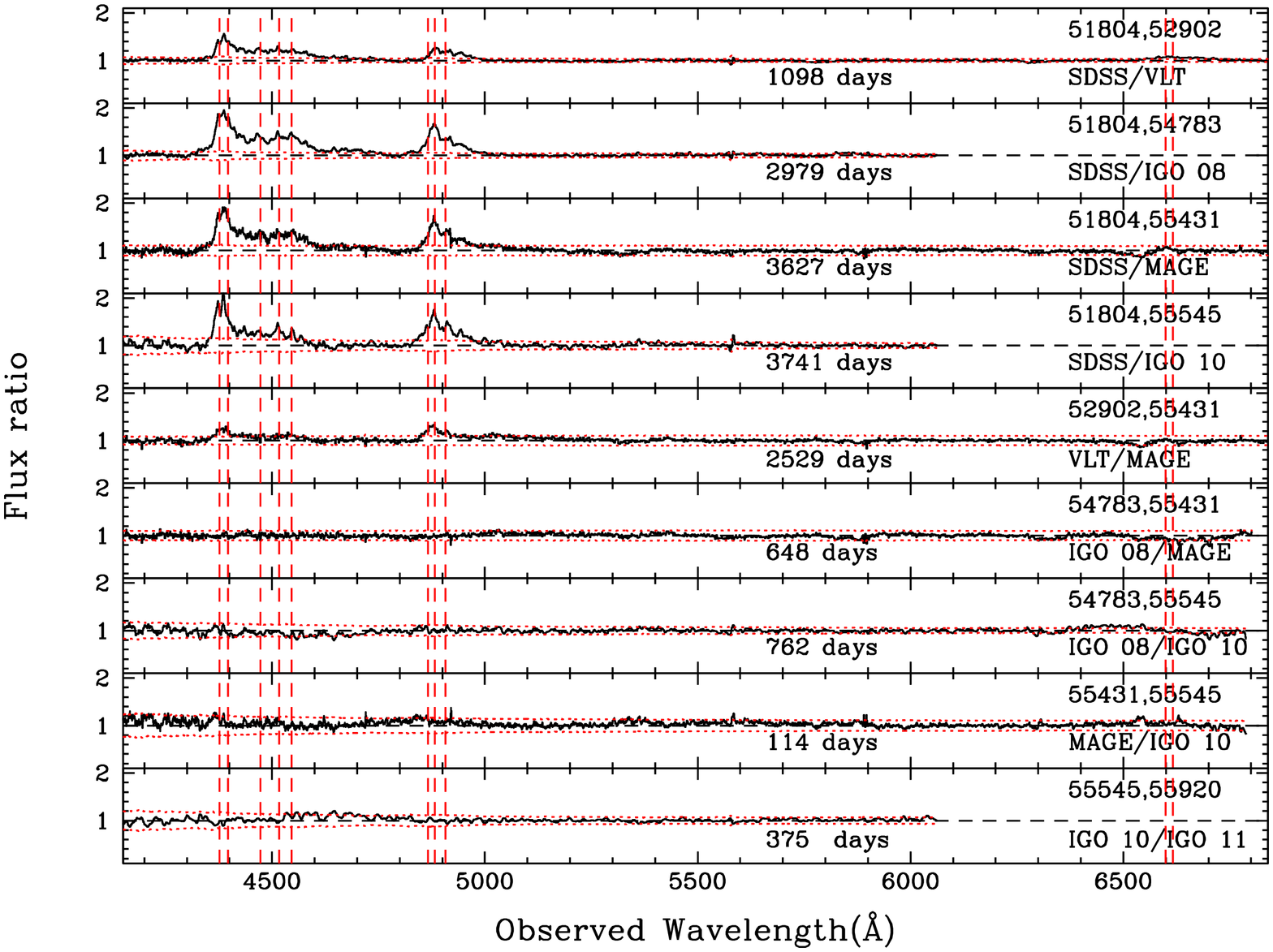,width=1.0\linewidth,height=0.7\linewidth,angle=270}%,bbllx=200bp,bblly=44bp,bburx=41bp,bbury=760bp}

\end{tabular}

 \caption{ SDSS, VLT, IGO and MagE spectra are plotted and labeled with the observation year in the top panel.  The absorption lines are marked with  dashed  red lines and the emission lines are marked with dotted  blue lines. The lower panels shows the ratio spectra between different epochs. The epochs of the ratio spectra and the difference between the corresponding MJD's  are labeled in each panel.} 
 \label{fig56}
 \end{figure*}
%+++++++++++++++++++++++++++++++++++++++++++++++++++++++++++++++++++++++++++++++++++++++++++++++++++++++++++++++++
%+++++++++++++++++++++++++++++++++++++++++++++++++++++++++++++++++++++++++++++++++++++++++++++++++++++++++++++++++

Fig.~\ref{fig56} shows the variability in the absorption lines of 
SDSS J2215-0045 between the different epoch data. In the top panels, 
we have overplotted the  SDSS, VLT, IGO and MagE data.   The plot 
clearly shows the variations in flux in the region of \aliii, \feiii\ UV 34, 
\feiii\ UV 48 and \mgii\ absorption lines.  In order to quantify the 
optical depth variability in Fig.~\ref{fig56} we plot ratio of spectra between 
different epochs together with the associated errors.
It can be seen that 
there is significant variability in \mgii, \aliii, \feiii\ UV 34 and 
\feiii\ UV 48 absorption lines. The absorption lines have maximum 
strength in the SDSS spectra. All the later epochs show decrease in the 
optical depth with respect to the SDSS data. 
The lines decreased in 
optical depth more significantly in the 2008 IGO spectra.
However, observations made after 2008 show that both MagE data and IGO data are consistent with
no variations in the absorption line optical depths.
Al~{\sc ii} 
absorption is not covered by our IGO spectra. It is clear from 
Fig.~\ref{J2215variable} that at the locations Al~{\sc iii}, 
\mgii\ and \feiii\ UV absorption lines, the observed spectra after the year 
2008 just follow the template.  This means that the data is consistent 
with  the disappearance of absorption seen in the SDSS spectrum. 
We also notice that the weak \alii\ absorption seen in SDSS spectrum 
has also disappeared in our MagE spectrum.   In Table~\ref{table3} we give
the average ratio measured between different epoch spectra at the
wavelength range covered by Fe~{\sc iii} UV 34 and 48 lines. If
there are no optical depth variation, the average ratio is 1.
This table confirms the visual trend seen in Fig.~\ref{fig56}.

\begin{table}
\caption{Average flux ratios in the region of absorption lines for  SDSS J2215$-$0045 }
\begin{tabular}{lcccc}
\hline
&&& \multicolumn{2}{c} {Average ratio }\\
Epoch1&Epoch2&$\Delta$t$^1$&Fe~{\sc iii}$^2$&Fe~{\sc iii}$^3$\\
&&&UV34&UV48\\
&&(days)&(\AA)&(\AA)\\
\hline
\hline
%
%205.989  7.083      94.298 4.863
%228.731  7.922     104.07  5.405
%221.028  7.654     103.978 5.399
%210.652  7.326     106.303 5.521
%188.107  10.224     95.264 7.460
%178.776  7.083      85.680 4.903
%168.521  6.686      86.839 4.969
%185.685  3.596      93.080 2.703
%189.982  7.529      86.954 4.958
%%
%
%
%
SDSS    &	VLT-FORS1  & 1098  & 1.14$\pm$0.04 &1.09$\pm$0.06\\
SDSS	&	IGO-2008   & 2979  & 1.27$\pm$0.04 &1.20$\pm$0.06\\
SDSS	&	MAGE       & 3627  & 1.23$\pm$0.04 &1.20$\pm$0.06\\
SDSS	&	IGO-2010   & 3741  & 1.17$\pm$0.04 &1.23$\pm$0.06\\
VLT-FORS1&	MAGE       & 2529  & 1.04$\pm$0.06 &1.10$\pm$0.09\\
IGO-2008  &	MAGE       &  648  & 0.99$\pm$0.04 &0.99$\pm$0.06\\
IGO-2008  &	IGO-2010   &  762  & 0.94$\pm$0.04 &1.00$\pm$0.06\\
MAGE	&	IGO-2010   &  114  & 1.03$\pm$0.02 &1.07$\pm$0.03\\
IGO-2010  &	IGO-2011   &  375  & 1.05$\pm$0.04 &1.00$\pm$0.06\\
\hline
\end{tabular}
\begin{flushleft}
$^1$ Elapsed time in observers frame; $^2$  Over the observed wavelength 
range 4316$-$4762\AA; $^3$ Over the observed wavelength range 4809$-$5024\AA.
\end{flushleft}
\label{table3}
\end{table}
 
\begin{figure}
  \centering
  \psfig{figure=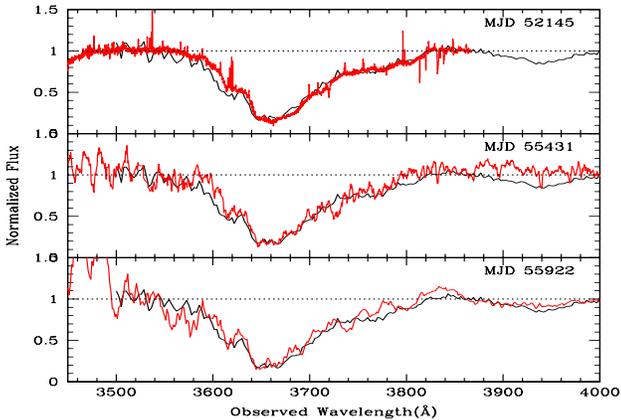,width=1.0\linewidth,height=0.7\linewidth,angle=0}
  \caption{ Normalized \civ\ absorption line, seen in  VLT/UVES, MagE and 
IGO/IFORS1 (in red/gray)  spectra, is compared to the VLT/FORS1 spectra 
obtained on MJD 52902 (black). In all these cases the continuum normalisation
is done using lower order polynomials connecting the absorption free regions
on either side of \civ\ absorption.}
%Continuum in all the cases are normalised using the available absorption free r%egions on either side of the \civ\ absorption.

 \label{figciv}
  \end{figure}

A strong \civ\ absorption is detected in the spectra covering 
$\lambda_{obs} < $ 4000 \AA.  Fig.~\ref{figciv} shows the normalized 
spectrum in the vicinity of  \civ\ absorption line which is covered in 
the VLT/FORS1, VLT/UVES, MagE and  the IGO/FORS1 data.  As \civ\ 
falls in a region which is free of Fe emission, we  fitted a low-order 
polynomial connecting the absorption free regions close to the \civ\ 
line for continuum normalization. We notice the \civ\ 
absorption is much broader than that of \aliii\ (or Al~{\sc ii}). 
This suggests that the two species may originate from two different 
regions (i.e., two distinct components of a co-spatial multiphase 
structure or from gas at distinctly different locations). The \siiv\ 
line falls at the edge of the MagE spectrum and is not covered by 
VLT/FORS1.  From the figure, it can be seen  that \civ\ has not 
changed significantly as the low ionization lines. It is also 
interesting to note the \civ\ line does not go to zero. This could 
either mean \civ\ is optically thin or the absorbing gas doesn't  
cover the background source.

In the following section we discuss the implications of the observed
variability in this source.

\section{Discussion on variability in SDSS J2215-0045}

Variability in broad absorption lines  has several origins. 
The simplest explanations are (1) change in the ionizing condition,
(2) proper motion of absorbing clouds \citep{ma02,hamann08,leighly09,krongold10,hidalgo11,vivek11} and (3) covering factor
variability of the absorbing gas \citep[see for example,][]{anand99}.
 
In the case of SDSS J2215$-$0045 the variability in the Al~{\sc iii}
and Fe~{\sc iii} fine-structure lines are unambiguously established.
As we do not have direct access to the column density of the
Fe~{\sc iii} ground state absorption, it will be difficult to draw
any conclusion on the  variations of the relative population 
of the excited level with respect to the ground level. However,
we know in the case of photoionization models Fe~{\sc iii} will closely
trace Al~{\sc iii}. Also the weakness of Al~{\sc ii} and the lack of
Fe~{\sc ii} absorption is consistent with the range in the ionization
parameter being very narrow \citep[see discussion on this by][]{dekool02}.
Thus any small change in the ionizing radiation will change both
Al~{\sc iii} and Fe~{\sc iii} column densities rapidly. Simple ionization
change (even without changing the excitation temperature) will change
the column density of Fe~{\sc iii} fine-structure levels. As discussed 
in Section 2, the light curve of SDSS J2215-0045 is consistent with the 
V-band photometric variability of $>$ 0.2 mag. The actual variability in 
the UV range  that controls the ionization state of the gas can be higher. 
From the Figs.~6 \& 7 of \citet{dekool02}, it is apparent that the ion 
fraction of \feiii\ (and \aliii) can be significantly reduced even for 
a small change in the ionization parameter. Compared to the SDSS epoch, 
our IGO data was taken when the QSO was 0.5 mag brighter. Therefore, 
the observed variability of \feiii\ and \aliii\ lines is consistent with 
photoionization induced variability.

The smallest timescale over which we have seen the variation
is between the SDSS in 2000 and VLT in 2003. This time period 
corresponds to a timescale of $\sim$ 1.211 years in the quasar 
rest frame. The variability time scale can be used to constrain the 
electron density of the absorbing gas. 
For a moderately ionized gas $n_{\rm i} > n_{\rm i+1}$, the recombination 
time scale can be approximated to \citep[see][]{anand01} 
\begin{equation}
t_r=(n_e \alpha_r)^{-1} .
\end{equation}
The variability time scale gives the recombination time scale, if the 
variations are assumed to be caused due to changes in optical depth. 
Using the recombination cross-section for \aliii\ given by the CHIANTI 
atomic database \citep{Dere97,dere09} the lower 
limit on the electron density, $n_e$, is given by
\begin{equation}
 n_{\rm e} \geq 1.3214 \times 10^4 t_{\rm yr}^{-1}
\end{equation}
where, $t_{\rm yr}$ is the elapsed time in years. The recombination 
coefficients are calculated for a temperature of 10,000 K.
Putting  $t_{\rm yr}=1.211$ in the above equation results in an 
electron density, $n_{\rm e} \geq 1.091\times 10^4$ cm$^{-3}$. This is 
coincident with but not as high as what has been suggested by the 
models of \citet{dekool02}.  However, the lack of variability in \civ\ 
under this scenario could mean that the absorbing gas has multiphase structure with \civ\ and \feiii\ absorption originating from different phases.

Models of FeLoBALs by \citet{faucher11} suggest the in situ formation of  
Fe BALs in the ISM of host galaxies by shocks induced by the QSO blast wave. 
Going along this line, \citet{rogerson11} put forward a model where an 
accelerating, relatively low density wind collides with dense \feiii\ 
clumps and produces the observed absorption lines. Since \feiii\ clumps 
are ablated in this process, the \feiii\ troughs will decrease with time 
unless  the wind encounters new clumps. The variability seen in the 
present system is consistent with this scenario. In this case, \feiii\ 
absorption is produced from a distinct region (probably the ISM of the 
host galaxy) compared to that of \civ. However, as this  time-scale 
should be shorter than sound crossing time for a T $\sim$10$^{4}$ K 
gas we can write the cloud thickness as,
\begin{equation}
 \Delta R = \frac{0.01\times t_{\rm yr}}{640} pc = 6\times10^{13} cm.
\end{equation}
This is much smaller than the size of the UV continuum emitting region. 
This thickness is much less than what one gets from photoionization models.

Recently several cases of emergence and subsequent disappearance of new 
components have been reported in the literature 
\citep{ma02,hamann08,krongold10,leighly09,hidalgo11,vivek11}. In all 
these cases, the absorbing gas transiting perpendicular to our line of 
sight is considered as a viable explanation. Such an explanation can 
hold for the present case also.

To investigate this scenario,  we derive some basic parameters for 
SDSS J2215-0045. The SDSS u and g magnitudes are converted in to B 
magnitude following the transformation equation of \citet{jester05} 
obtained for $z\le2.1$ QSOs.  We get the bolometric luminosity for this 
source from the B magnitude as 2.7$\times$10$^{47}$ erg s$^{-1}$ using the 
prescription of \citet{marconi04}.  The bolometric luminosity corresponds 
to a black hole mass of 2.2$\times$10$^9 M_{\odot}$. The diameter of the 
disc within which 90\% of the 2700 \AA$~$ continuum is emitted is obtained 
as 1.49$\times$ 10$^{16}$ cm ($\sim$ 46 times the Schwarzschild radius)
as expected in a Shakura \& Sunyaev thin accretion disk 
model \citep{hall11}.  Variability in SDSS J2215-0045 has occurred between 
the SDSS and IGO observations. If we take the number of days elapsed 
between the SDSS and IGO observations (3740 days in observer's frame 
or 1509 days in quasar frame) as an upper limit on the transit time of the 
gas, the transverse velocity of the gas can be estimated as 
$v_\perp\ge$ 1140 \kms\ if we assume the projected size of the gas to be 
much smaller than the emitting region and a face on disk. Based on the 
redshift of the \aliii\ absorption  we infer the line of sight velocity 
to be of the order of 15,000 \kms. Therefore, like in the case of 
J1333+0012 \citep{vivek11} the outflow should be very close to the line of 
sight such that the transverse  velocity is much smaller than the line of 
sight velocity. 
% The peak of the \mgii absorption corresponds to a typical line of velocity $\sim$ 15000\kms.

%{\bf Anand: Actually
%the last statement needs to be checked as we notice the strong variability
%prior to 2008. So it will be important to get data prior to this epoch. 
%Ashish it will be great if you can get the light curve for the fifth
%source and also extend the data to earlier epoch specially for 2215.}
%It is clear from the previous plots that the \mgii,\aliii and fine structure \feiii absorptions lines have undergone a significant variability between the years 2000 and 2010. \aliii and \mgii components are seen to have completely vanished in our latest observations. 
\section{Results \& Discussion}

We present the results of spectroscopic monitoring of 5 FeLoBALs 
for a period of 10 years. We also present the photometric light 
curves of all these sources obtained from Catalina Real-Time 
Transient Survey. 

In one of these QSOs, SDSS J2215$-$0045, we detected the 
absorption line variability of Al~{\sc iii} and Fe~{\sc iii}
fine-structure lines. However, there is no clear variation
in the absorption profile of C~{\sc iv} absorption. Our
results are consistent with low-ions and C~{\sc iv} originating
from different components along the line of sight. The 
absorption line variability could be related to changes
in the ionization state of the gas and/or to changes in the covering
factor due to transverse motion of the gas. The
light curve of SDSS J2215$-$0045 suggests brightening of
this QSO when the absorption line became weak. This
together with the expected narrow range in the allowed
parameters of the photoionization models \citep{dekool02,rogerson11}
means the observed variability can very well be explained
by changes in the photoionization rates. The data is also
consistent with the models of \citet{rogerson11} in which a shock 
heated cloud that produces Fe~{\sc iii} absorption being ablated
producing strong variations in the absorption strength. 
Regular photometric and spectroscopic monitoring of this source
is needed to distinguish between these alternatives.

In the remaining cases no significant variation in the 
absorption line is detected. As these sources do not show 
any strong flux variability photoionization induced absorption 
line variability is not expected
in these sources. However, if the low ion absorption in 
these systems are due to ISM (or high density clump 
far away from the central QSO) being shock heated by
QSO feedback then we do not find any evidence of
this gas being ablated with time.   Also, we can conclude
that either the projected extent of the gas is larger than the continuum emitting region or the transverse velocity is small.

If FeLoBALs are different set of population (where the
absorption occurs far away from the the central engine), then 
comparing the occurrence of variability in these objects with those of HiBALs and LoBALs without fine-structure lines will throw light on the nature of BAL phenomenon in QSOs.

\section{acknowledgements}
{We wish to acknowledge the IUCAA/IGO staff for their support during our observations and Maria-Jose Maureira for help with MagE observations. We 
also thank Dr. DiPompeo for sharing their VLT/FORS data with us and the anonymous referee for useful suggestions.
{MV gratefully acknowledges University
Grants Commission, India, for financial support through RFSMS Scheme
and IUCAA for hospitality, where most of this work was done.
}
RS and PPJ gratefully acknowledge support from the Indo-French
Centre for the Promotion of Advanced Research (Centre Franco-Indien pour
la promotion de la recherche avanc\'ee) under Project N.4304-2.

}

%----------------- Bibliography and bibfile ------------------------------------------
\def\aj{AJ}%
\def\actaa{Acta Astron.}%
\def\araa{ARA\&A}%
\def\apj{ApJ}%
\def\apjl{ApJ}%
\def\apjs{ApJS}%
\def\ao{Appl.~Opt.}%
\def\apss{Ap\&SS}%
\def\aap{A\&A}%
\def\aapr{A\&A~Rev.}%
\def\aaps{A\&AS}%
\def\azh{AZh}%
\def\baas{BAAS}%
\def\bac{Bull. astr. Inst. Czechosl.}%
\def\caa{Chinese Astron. Astrophys.}%
\def\cjaa{Chinese J. Astron. Astrophys.}%
\def\icarus{Icarus}%
\def\jcap{J. Cosmology Astropart. Phys.}%
\def\jrasc{JRASC}%
\def\mnras{MNRAS}%
\def\memras{MmRAS}%
\def\na{New A}%
\def\nar{New A Rev.}%
\def\pasa{PASA}%
\def\pra{Phys.~Rev.~A}%
\def\prb{Phys.~Rev.~B}%
\def\prc{Phys.~Rev.~C}%
\def\prd{Phys.~Rev.~D}%
\def\pre{Phys.~Rev.~E}%
\def\prl{Phys.~Rev.~Lett.}%
\def\pasp{PASP}%
\def\pasj{PASJ}%
\def\qjras{QJRAS}%2215.bib
\def\rmxaa{Rev. Mexicana Astron. Astrofis.}%
\def\skytel{S\&T}%
\def\solphys{Sol.~Phys.}%
\def\sovast{Soviet~Ast.}%
\def\ssr{Space~Sci.~Rev.}%
\def\zap{ZAp}%
\def\nat{Nature}%
\def\iaucirc{IAU~Circ.}%
\def\aplett{Astrophys.~Lett.}%
\def\apspr{Astrophys.~Space~Phys.~Res.}%
\def\bain{Bull.~Astron.~Inst.~Netherlands}%
\def\fcp{Fund.~Cosmic~Phys.}%
\def\gca{Geochim.~Cosmochim.~Acta}%
\def\grl{Geophys.~Res.~Lett.}%
\def\jcp{J.~Chem.~Phys.}%
\def\jgr{J.~Geophys.~Res.}%
\def\jqsrt{J.~Quant.~Spec.~Radiat.~Transf.}%
\def\memsai{Mem.~Soc.~Astron.~Italiana}%
\def\nphysa{Nucl.~Phys.~A}%
\def\physrep{Phys.~Rep.}%
\def\physscr{Phys.~Scr}%
\def\planss{Planet.~Space~Sci.}%
\def\procspie{Proc.~SPIE}%
\let\astap=\aap
\let\apjlett=\apjl
\let\apjsupp=\apjs
\let\applopt=\ao
\bibliographystyle{mn2e}
\bibliography{2215}
%\label{lastpage}
\end{document}